\documentstyle[12pt]{article}
\input epsf

\def\hc{{\dagger}} 
\def\tp{{\scriptscriptstyle T}} 
\def\tr{\hbox{tr}} 
\def\422{{$SU(4)\otimes SU(2)_L\otimes SU(2)_R$}}
\def\diag.{\hbox{diag.}}
\def\muegamma{\hbox{$\mu\to\hbox{e}+\gamma$\ }}
\def\taumugamma{\hbox{$\tau\to\mu+\gamma$\ }}
\def\refeqn#1{(\ref{#1})}
\def\MGUT{\hbox{$M_{PS}$\ }}
\def\SM{{SM}}
\def\tanb{\hbox{$\tan \beta$}}
\def\GeV{{GeV}}
\def\MSSM+N{\hbox{MSSM+$\nu$}}
\def\etal{{\it et al.}}

\def\FigLambdaTop{1}
\def\FigSusyDiag{2}
\def\FigUll{3}
\def\FigAr{4}
\def\FigDelta{5}
\def\FigJ{6}
\def\FigSleptons{7}
\def\FigSneutrinos{8}
\def\FigNeutralinos{9}
\def\FigLighestScalar{10}
\def\FigBRmuegammaMain{11}
\def\FigBRmuegammaMnu{12}
\def\FigBRmuegammaAo{13}
\def\FigBRtaumugammaMain{14}

\def\FigNewBRmuegammmaMain{15}
\def\FigNewBRtaumugammmaMain{16}
\def\FigNewNewBRtaumugammma{17}
\def\FigNewSleptons{18}
\def\FigNewLightestScalar{19}
\def\FigNewSneutrinos{20}
\def\FigNewLightestCharginoNeutralino{21}
\def\FigNewCharginoNeutralinoB{22}

\begin{document}
\baselineskip 24pt
\newcommand{\sheptitle}
{Lepton Flavour Violation in String-Inspired Models}

\newcommand{\shepauthor}
{S. F. King$^{\dagger}$
\footnote{On leave of absence from $^\ast$.}
and 
M.Oliveira$^\ast $
\footnote{Work supported by JNICT 
under contract grant : PRAXIS XXI/BD/5536/95.}}

\newcommand{\shepaddress}
{$^\dagger$ Theory Division, CERN, CH-1211 Geneva 23, Switzerland\\
$^\ast$Department of Physics and Astronomy, University of Southampton \\
        Southampton, SO17 1BJ, U.K}

\newcommand{\shepabstract}
{Lepton flavour violation (LFV) has been proposed as a significant
test of supersymmetric unification. Here we show that such signals
are also a generic feature of supersymmetric string unified models
in which there is no simple unified gauge group. In realistic
models of this kind which involve third family Yukawa unification and
large values of \tanb, there are generally
heavy right-handed (singlet) neutrinos of intermediate mass
$M_{\nu}$, whose couplings violate lepton flavour.
To illustrate these effects we calculate the rates for
\muegamma and \taumugamma in the minimal 
supersymmetric \422 model.
Including only the minimum irreducible
contributions, we find that
both rates are enhanced relative
to similar models with low $\tan \beta$, with
\taumugamma providing a decisive test of such models
in the near future.}

\begin{titlepage}
\begin{flushright}
CERN-TH/98-28\\
SHEP-98/05\\
hep-ph/9804283\\
\end{flushright}
\begin{center}
{\large{\bf \sheptitle}}
\\ \shepauthor \\ \mbox{} \\ {\it \shepaddress} \\ 
{\bf Abstract} \bigskip \end{center} \setcounter{page}{0}
\shepabstract
\begin{flushleft}
CERN-TH/98-28\\
\today
\end{flushleft}
\end{titlepage}


\section{Introduction.}

Recently there has been much interest in 
lepton flavour violation (LFV) as a probe of
physics beyond the standard model triggered by the observation of
Barbieri and Hall \cite{bh, barb&hall} that processes such as
\muegamma might be very good low energy signals
of supersymmetric grand unified theories (SUSY GUTs). 
In the standard model separate lepton numbers $L_e,L_{\mu},L_{\tau}$
are exactly conserved, which explains the absence of LFV to 
remarkable accuracy (the present limit on the branching ratio for
\muegamma is approaching $10^{-11}$.) Even if small neutrino
masses are introduced into the standard model, thereby violating separate
lepton numbers, the effect on \muegamma is very small,
since the amplitude is proportional to $\Delta m_{\nu}^2/M_W^4$
multipled by suitable mixing angles, where $\Delta m_{\nu}^2$ is the
difference in the squared masses of two neutrino species, and
$M_W$ is the W boson mass. The introduction of SUSY allows the possibility
of larger contributions to such processes since the soft SUSY breaking 
masses and couplings may violate separate lepton numbers by arbitrarily
large amounts.
This means that in SUSY there are in general additional diagrams
which have in principle large contributions to LFV processes
\cite{SUSYLFV}.

One way to avoid conflict with the experimental limits is to
invoke some supergravity (SUGRA)
theory \cite{SUGRA} which leads to universal soft parameters 
at the Planck scale \cite{review}. In the absence of radiative corrections
the selectron mass matrix in the basis $\tilde{e},\tilde{e^c}^\ast$
would look like
\begin{equation}
\left(\begin{array}{cc}
m_e^\dagger m_e +m_{3/2}^2I  & Am_e \\   
A^\ast m_e^\dagger  & m_e m_e^\dagger+m_{3/2}^2I      
\end{array} \right) \label{selectronmatrix}
\end{equation}
where $m_e$ is the electron mass matrix, I is the unit matrix,
and $m_{3/2}$, $A$ are universal soft parameters.
Clearly each $3\times 3$ block of
the selectron mass matrix becomes diagonal in the basis in which
the electron mass matrix is diagonal, which implies no LFV.
Any violation of universality
will lead to off-diagonal elements
in the $3\times 3$ blocks of the slepton matrix in the charged lepton
mass basis, which implies LFV.
In the minimal supersymmetric standard
model (MSSM) this result is preserved even in the presence of radiative 
corrections.
This is because the renormalisation group (RG)
equations do not generate any off-diagonal elements for squark masses
in a basis in which the charged lepton Yukawa couplings are diagonal.
Clearly if the universality assumption is relaxed then arbitrarily large
LFV is possible in the MSSM. The observation of Barbieri and Hall is that
with SUSY GUTs LFV is unavoidable, even with the assumption 
of universality. Part of the reason is that in GUTs,
quarks and leptons share a common multiplet so that
the lepton sector is contaminated by the
flavour violating quarks. Without SUSY such an effect, though present, 
would be generally very weak as it scales 
with an inverse power of the scale of the 
unification scale $M_{GUT}$. 
However in the presence of SUSY the RG running of the slepton masses
between $M_P$ and $M_{GUT}$ causes the LFV to be imprinted onto the
slepton masses, which are no longer diagonal in the basis in which the 
leptons are diagonal. Below the GUT scale the MSSM RGEs then ensure that
the LFV effect is preserved down to the TeV scale
where it may lead to sizeable contributions to physical processes.

Although LFV can be interpreted as a signal of SUSY GUTs
such as $SU(5)$, $SO(10)$
\cite{bh, barb&hall} similar effects
can be achieved without the presence of a GUT gauge group,
even assuming strictly universal soft parameters at $M_P$.
For example, simply adding a 
right-handed neutrino to the MSSM
(\MSSM+N) \cite{hisano} will generate LFV effects due to the
fact neutrino that (Dirac) Yukawa couplings are not diagonal in the
basis in which the charged lepton Yukawa couplings are diagonal.
In the charged lepton mass basis
the non-diagonal neutrino Yukawa matrix will generate off-diagonal 
slepton masses 
due to the RG running of the slepton mass matrix
between $M_P$ and the scale $M_\nu$, where $M_\nu$
is the Majorana mass scale of the right-handed neutrinos.
This will result in low energy LFV effects rather similar to those
in SUSY GUTs, but without any underlying GUT gauge group.
However the \MSSM+N theory is rather unconstrained compared to
SUSY GUTs, and it is of interest to see if similar effects could
occur in other better motivated, but non-GUT models. In particular
we have in mind string-inspired models which do not involve a simple
gauge group, but where the gauge couplings are unified at the string scale.

In this paper we shall focus on a particular 
recently proposed string-inspired model, the
minimal supersymmetric \422 model \cite{shafi}
(see also \cite{pati, antoniadis}). In this model,
quarks and leptons are unified into common multiplets, but there is
no simple GUT gauge group. Instead the gauge couplings are unified with
gravity at the string scale. In the minimal version
\cite{shafi} the only source of LFV in the 422 model is via the
right-handed neutrino couplings, as in the \MSSM+N model.
However, unlike the \MSSM+N model, the minimal 422 model is much more
highly constrained. For example in minimal 422 there is complete Yukawa
unification for the third family top, bottom, tau and tau-neutrino Yukawa
couplings, which automatically leads to the prediction of a 
large ratio of Higgs vacuum expectation values (VEVs) with
\tanb\ in the range 30-60 \cite{ben&king459} which lies
beyond the scope of the results presented in \cite{hisano}.
As discussed in \cite{arkani} the large \tanb\ region involves
some new effects which were neglected in previous treatments,
and we are careful to include all relevant effects here.

It is worth comparing the minimal string inspired
\422 model to $SO(10)$, which also may have
Yukawa unification. In $SO(10)$ colour triplets with couplings
to fermions are inevitably present in the effective theory
beneath $M_P$. Indeed in $SU(5)$ this is the primary source of LFV.
But in minimal 422 such colour triplets, although genericaly present,
do not couple to fermions, and play no role in LFV. \footnote{
In a non-minimal 422 model this source of flavour violation could be
included at the cost of introducing eighteen unconstrained new
parameters. } In general it is
easy to introduce new sources of LFV, for example via LFV soft mass terms
which for example may be controlled by
additional $U(1)_X$ gauged family symmetries
\cite{softleo}. Our approach here is to consider the {\em minimum
irreducible} amount of LFV associated with this class of model.
Such an approach allows unavoidable constraints to be placed on the model 
from the experimental limits on LFV.

Concerning our results, we find that
the diagrams involving sneutrinos and charginos in the loop
are found to give the dominant contribution to \muegamma and \taumugamma .
The off-diagonal sneutrino masses, which are responsible for LFV,
receive contributions from two distinct sources:
F-term Dirac neutrino masses (which always occur at the electroweak scale
despite the fact that the right-handed neutrinos are much heavier)
and RGE evolution in the high energy region between the Planck scale
and the right-handed neutrino mass scale.
Although both effects result from the neutrino Yukawa couplings,
they enter with opposite sign, and can lead to cancellations
in some regions of parameter space. Nevertheless in the minimal 
supersymmetric \422 model, with realistic masses and mixing angles,
we find that generally that the rates are enhanced relative to the
low \tanb\ case due to increased 12 and 23 family mixing effects. 
In particular we find that the predicted rate for \taumugamma 
is quite close to the current experimental limit.

The organisation of the rest of the 
paper is as follows. In Section 2
we review the minimal 422 model. In Section 3 we
describe in detail how the model was implemented giving particular
emphasis to boundary conditions of the RGEs. Section 4
is devoted to a detailed analysis of the one-loop decay \muegamma
and \taumugamma.
Section 5 contains our conclusions.


\section{The Model.}

Above \MGUT we have adopted a model with unified gauge group \cite{pati}
\begin{equation}
G_{PS}=SU(4)\otimes SU(2)_L \otimes SU(2)_R 
\end{equation}
Here we briefly summarise
the parts that are relevant for our analysis.
For a more complete discussion see \cite{antoniadis}. 
The left-handed quarks and leptons are accommodated 
in the following $F=(4,2,1)$, $F^c=(\bar 4,1,\bar 2)$ representations :
\begin{equation}
F_i = \left(
                 \matrix{
                 u^r & u^b & u^g & \nu \cr
                 d^r & d^b & d^g & e \cr}
                 \right)_i
\qquad
F_j^c = \left(
                      \matrix{
                      d^c_r & d^c_b & d^c_g & e^c \cr
                      u^c_r & u^c_b & u^c_g & \nu^c \cr}
                      \right)_j
\end{equation}
where $i,j=1\ldots 3$ is a family index. 
The MSSM Higgs fields are contained in $h=(1,\bar 2,2)$ :

\begin{equation}
h=
\left(\begin{array}{cc}
H_u^0 & H_d^+ \\   
H_u^- & H_d^0      
\end{array} \right) \label{h:higgs}
\end{equation}
\noindent
whereas the heavy Higgs $H=(4,1,2)$ and $H^c=(\bar 4,1,\bar 2)$ are denoted:
\begin{equation}
H = \left(
                  \matrix{
                  H_{u^r} & H_{u^b} & H_{u^g} & H_{\nu} \cr
                  H_{d^r} & H_{d^b} & H_{d^g} & H_{e} \cr}
                  \right)
\qquad
H^c= \left(
                     \matrix{
                     H_{u^c_r} & H_{u^c_b} & H_{u^c_g} & H_{\nu^c} \cr
                     H_{d^c_r} & H_{d^c_b} & H_{d^c_g} & H_{e^c} \cr}
                     \right) \label{H:higgs}
\end{equation}
\noindent
In addition to the Higgs fields in \refeqn{h:higgs} and
\refeqn{H:higgs} the model also involves an $SU(4)$ sextet 
field $D=(6,1,1)=(D_3,D_3^c)$. 

The superpotential of the minimal 422
model is \cite{shafi}:
\begin{eqnarray}
{\cal W} & = & S[\kappa ({{H}^c} H  - M_{PS}^2) +  \lambda  h^2]
+ \lambda_HDHH + {\lambda}_{H^c}D{H^c}{H^c} \nonumber \\
& + &  \lambda_{33} {F^c}_3 F_3 h 
+\lambda_{ij} {F^c}_iF_j h\frac{({H^c} H)^n}{M_P^{2n}} 
+ \lambda_{\nu ij} \frac{{F^c}_i {F^c}_j H H}{M_P} 
\label{susypot}
\end{eqnarray}
where $S$ denotes a gauge singlet superfield, the parameters $\kappa, 
\lambda$ and $M_{PS}$ are taken to be real and positive, and $h^2$ denotes 
the unique bilinear invariant $\epsilon^{ij} h^{(1)}_i h^{(2)}_j$.  
Also, $M_P (\simeq 2.4 \times 10^{18}$ GeV) denotes the `reduced' Planck 
mass.  As a result of the superpotential terms involving the singlet $S$ the
Higgs fields develop VEVs,
$\langle H \rangle = \langle H_\nu \rangle \sim M_{PS}$ and
$\langle H^c \rangle = \langle H_{\nu^c} \rangle \sim M_{PS}$,
which lead to symmetry breaking
\begin{equation}
SU(4)\otimes SU(2)_L\otimes SU(2)_R\to
SU(3)_c\otimes SU(2)_L\otimes U(1)_Y. \label{422:321}
\end{equation}
The singlet $S$ itself also naturally develops a small VEV of order
the SUSY breaking scale \cite{shafi} so that
the $\lambda S$ term in \refeqn{susypot} gives an effective 
$\mu$ parameter of the correct order of magnitude.
Under \refeqn{422:321} the Higgs field $h$ in \refeqn{h:higgs}
splits into the familiar MSSM doublets 
$H_u$ and $H_d$ whose neutral components subsequently develop weak 
scale VEVs $H_u^0=\langle v_u \rangle$ and
$H_d^0=\langle v_d \rangle$ with $\tanb = v_u/v_d$.

This model has Yukawa unification for the
third family \cite{arason&castano,bbo} which leads to a 
large top mass $m_{top} > 165$ \GeV\ and
$\tanb \sim m_{top}/m_{bottom}$. First and second family
Yukawa couplings are effectively generated by non-renormalisable
operators which are suppressed by powers of a heavy scale $M > M_{GUT}$. 
In the 422 model, these operators can be constructed from 
different theoretical group contractions of the fields 
such as \cite{king}:
\begin{equation}
{\cal O}_{ij} = {F_i}^{c} \lambda_{ij} F_j h 
                \left( H H^c \over M_P^2 \right)+\hbox{h.c.} \label{operator}
\end{equation}
The idea is that when the heavy Higgs develop their large
VEVs such operators
reduce to effective Yukawa couplings of the form $F^c\lambda F$ with a small
$(M^2_{PS}/M_P^2)$ coefficient. Assuming a (well motivated) 
texture \cite{texture} for the Yukawa matrix at \MGUT and suitably choosing a
set of operators, successful predictions can be made for some \SM\ 
parameters. Vertical splittings within a particular family are accounted
for by group theoretical Clebsch factors \cite{king}.
A detailed analysis of this approach for the 422 model
can be found in \cite{ben&king456}. The non-renormalisable
operators involving the right-handed neutrino result in Majorana masses
of the form 
$1/2\>M_\nu \nu^c \nu^c$, where $M_\nu \sim M^2_{PS}/M_P$,  
which enables right-handed neutrinos to decouple 
at the scale $M_\nu$, leading to a Gell-Mann-Ramond-Slansky
see-saw mechanism.

The $D$ field doesn't develop a VEV but the terms $HHD$ and 
$H^c H^c D$ combine the colour triplets parts of $H$, $H^c$
and $D$ into acceptable GUT scale mass terms \cite{antoniadis}. 
We note that the 422 symmetry also allows the couplings :
\begin{eqnarray}
FFD       &\to & Q Q D_3 + Q L D_3^c \\
F^c F^c D &\to & u^c d^c D_3^c + u^c e^c D_3 + d^c \nu^c D_3
\end{eqnarray}
which obviously would generate additional LFV signals. However
these may be forbidden by a global R-symmetry \cite{shafi}.
Their exclusion here is in keeping with the general philosophy of the approach
which is to consider the minimum amount of irremovable LFV in the model,
so that LFV becomes an unavoidable signal of the model.


\section{Procedure.}

In this section we describe how the 
422 model was implemented.
We considered three fundamental scales : $M_{SUSY}$ which was 
assumed to equal the top mass, $M_{GUT}\sim 2\times 10^{16}$ \GeV\ the scale 
of coupling unification and $M_{Planck}\sim 2\times 10^{18}$ \GeV. 
An additional scale $M_{\nu}$ describing the energy at which 
the right-handed neutrinos decouple via see-saw mechanism was
introduced. 
Cosmological constraints require 
$10^{10}$ \GeV\ $\le M_{\nu} \le M_{GUT}$ \cite{ben&king353}. 
Experimentally viable boundary
conditions were imposed at each fundamental scale 
and we used one-loop matrix RGEs \cite{martin&vaughn}
to relate parameters at different energies 
(particle threshold effects were ignored). 

We now turn to describe the algorithm of the program. Since we want to
achieve third family Yukawa unification (3FYU) at $M_{GUT}$
\cite{ben&king328}, 
which crucially depends on the unknown low energy values of 
$m_{top}$, $m_{\nu_\tau}=v_u \lambda_{\nu_\tau}$ and $\tan \beta$ 
a iterative procedure is needed. Initial estimates for these
parameters are guessed and, along with gauge couplings and all Yukawa
matrices, are run, first from their definition values to
$M_{SUSY}=m_{top}$ and afterwards from $M_{SUSY}$ to $M_{GUT}$.
At $M_{GUT}$ the guesses are tested to see if they actually
lead to 3FYU. This is unlikely to happen
in the first attempt therefore we induce slight changes in our
initial guesses and repeat the above process again. 
After a few iterations it starts to be obvious that some guesses are 
more sucessful than others. These are subjected to further
pertubative analysis allowing more precise GUT unification. 
This recursive approach is repeated many times
until the condition
$\lambda_{t}=\lambda_b=\lambda_{\nu_\tau}=\lambda_\tau$
is verified to a satisfactory accuracy (typicaly 1\%).
At this point some comments are worth making. 
The running of masses to $M_{SUSY}$ was done using Standard Model RGE 
(one-loop QED, three-loop QCD \cite{gorshny&tarasov}) 
and it is necessary as it provides important mass
corrections especially for light quarks. After the whole iterative
process is complete we are left with predictions for our guesses based
on the assumption of 3FYU. To have an idea, for the input $\alpha_s=0.115$,
$m_b=4.25$ \GeV, $M_{\nu}=M_{GUT}$ we obtained the following 
values $m_{top}\sim 175$ \GeV, $m_{\nu_\tau}\sim 122$ \GeV\ 
and $\tan\beta\sim 56$.
The dependence on $M_{\nu}$ is mostly felt by $m_{\nu_\tau}$ 
which can decrease to $115$ \GeV\ when $M_\nu\sim 10^{12}$ \GeV. 
The variations with
$\alpha_s$ and $m_b$ are considerable and well documented
\cite{ben&king456}. We note that
it is important to watch for the magnitude of the third family of
the Yukawa couplings at $M_{Planck}$ because they are very 
sensitive to $\alpha_s$ and $M_\nu$, 
therefore can easily acquire values outside
the range allowed in perturbative regime (Figure \FigLambdaTop).

\vbox{
\noindent
\hfil
\vbox{
\epsfxsize=8cm
\epsfysize=8cm
\epsffile[160 460 580 720]{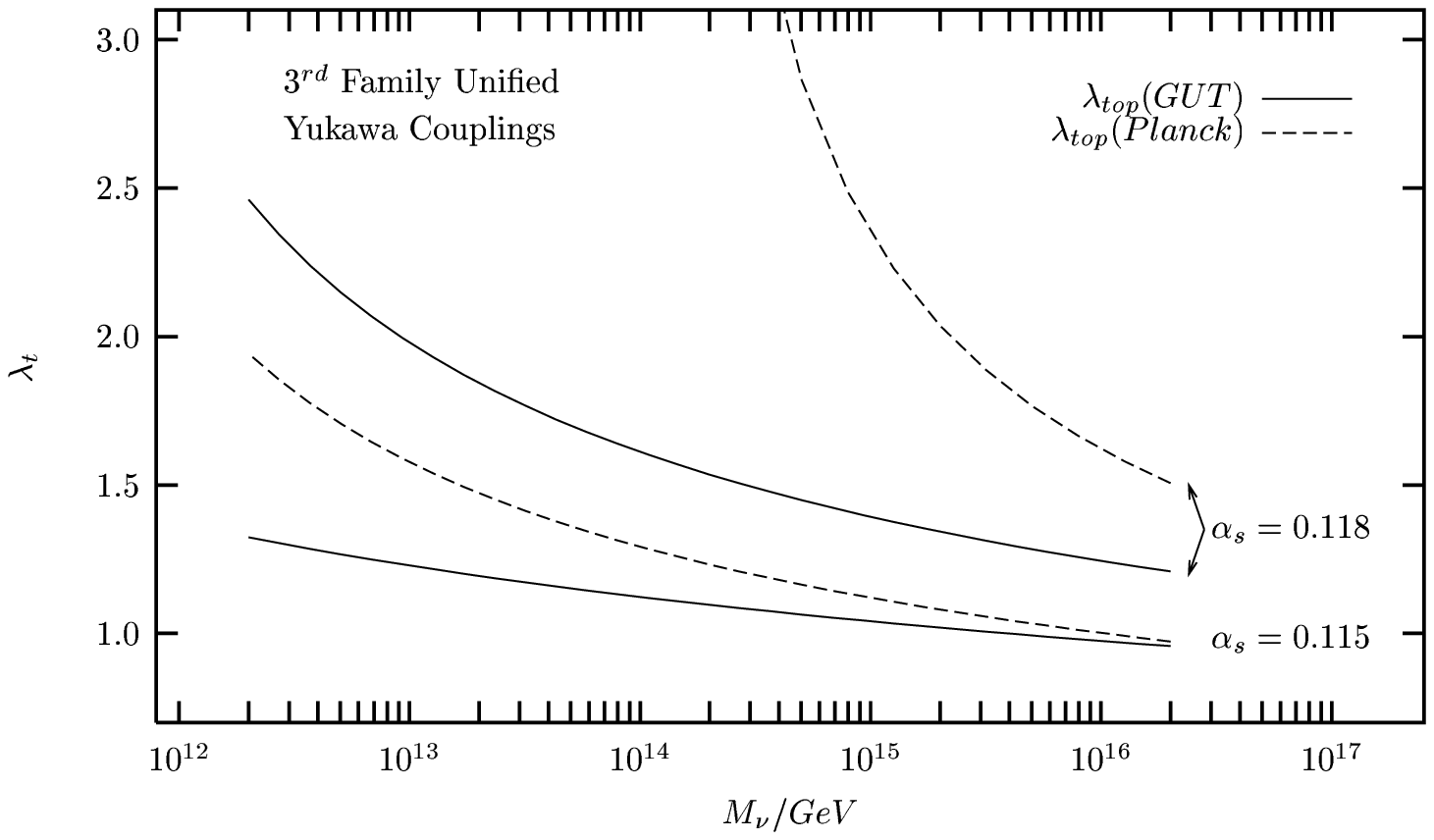}}

{\narrower\narrower\footnotesize\noindent
{\bf Figure \FigLambdaTop.} Dependence of third family unified Yukawa coupling
$\lambda_{top}$ at GUT (solid) and Planck (dashed) energy 
with the right-handed neutrino decoupling scale $M_\nu$ 
for two values of $\alpha_s$.\par
\bigskip}}

At $M_{PS}$, we must match all the Yukawa matrix couplings
to the ones which can be obtained from non-renormalized operators like
Eq.\refeqn{operator}. These new Yukawa matrices are not unique. 
However they are constrained
by the fact that they must predict the same (known) physics as the ones
they replace, i.e. both must have identical eigenvalues and quark mixing
angles (for the sake of simplicity we did not considered CP
violating phase).
Several forms of these Yukawa matrices were extensively studied in
\cite{ben&king456,ben&king459} for the 422 model. 
Here we will only consider the following particular one~
\footnote{Other choices would lead to same order of magnitude results.}~:

\begin{eqnarray}
\lambda_u   &=& \left( \matrix{  0    &  Y^{n=3} & 0      \cr
                               Y^{Ad} &  Y^D-Y^C & 0      \cr
                               0      &  Y^B     & Y^{33} \cr} 
                               \right) \label{up:opr} \\
\lambda_d   &=& \left( \matrix{   0      &   Y^1      & 0      \cr
                               3\ Y^{Ad} & -(Y^D+Y^C) & 0      \cr
                                  0      & - Y^B      & Y^{33} \cr}
                             \right) \\
\lambda_e &=& \left( \matrix{ 0           &  Y^1          & 0      \cr
                              9/4\ Y^{Ad} & 3\ (Y^D+Y^C)  & 0      \cr
                              0           & - Y^B         & Y^{33} \cr}
                            \right) \\
\lambda_\nu &=& \left( \matrix{  0         &    Y^{n=3}    & 0      \cr
                               3/4\ Y^{Ad} & -3\ (Y^D-Y^C) & 0      \cr
                                 0         &   Y^B         & Y^{33} \cr}
                               \right) \label{nu:opr} \\
\nonumber
\end{eqnarray}
\medskip
\noindent
We briefly explain their form. The zeros in positions $31$, $13$ are
motivated by correspondingly small entries on quark CKM matrix. The
zero in $23$ only effects the right-handed 
mixing matrix (because of high family hierarchy), 
thus it is convenient as it improves predictability. Two operators
were needed in the $22$ position because of particular high charm-muon
splitting \cite{ben&king456}. 
The operator $Y^B$ generates $V_{23}\sim V_{32}$, while
$Y^{Ad}$ and $Y^1$ generate $V_{12}\sim V_{21}$ and first family
masses.\footnote{Actually, since the ($n=2$) $Y^1$ operator has a
vanishing Clebsh for the up-type fermions, we are forced to introduce
a further $Y^{n=3}$ operator in the 12 position, if we want to avoid a
massless up quark.} The coefficients on different matrices 
associated with the same operator $Y$ are the Clebsh 422 factors mentioned
in Section 2.
Solutions for $Y$s were numerically searched for 
the input: $V_{23}$, $m_{charm}$, $m_\mu$ and
$V_{12}$, $m_{up}$, $m_e$, which enabled three quantities to be
predicted~:
$V_{13}$, $m_{down}$, $m_{strange}$ (see Appendix 4 for results). 
Notice that this model predicts
the experimentally unavailable Dirac neutrino 
masses $m_{\nu_e}$,$m_{\nu_\mu}$,$m_{\nu_\tau}$ and the 
lepton Dirac CKM matrix $V^L$. 
The prediction of physical neutrino masses
and mixing angles relies on knowledge of the right-handed neutrino
Majorana mass matrix $M_{\nu}$. Following \cite{ben&king459} we shall
assume that $M_{\nu}$ is proportional to the unit matrix.
This rather {\it ad hoc} assumption at least
has the virtue that it leads the result
that the modulus of the
leptonic CKM matrix elements are equal to those calculated just from
the Dirac neutrino mass parts \cite{ben&king459}. It also means that
the physical neutrino masses are determined by a single mass parameter
which we continue to denote by $M_{\nu}$, where this parameter 
henceforth refers
to the overall factor multiplying the unit Majorana matrix rather than
the matrix itself. With a suitable choice of $M_{\nu}$ 
this simple assumption leads to a physical muon-neutrino and electron-neutrino
mass spectrum suitable for the MSW solution to the solar neutrino 
problem, with a tau-neutrino in the correct range for hot
dark matter, and with muon-tau neutrino oscillations in the observable range
of the CHORUS experiment \cite{ben&king459}. If this assumption is relaxed
one would generally expect qualitatively similar effects both in the 
neutrino spectrum, and in the physics of LFV considered here.

After having set experimentally viable Yukawa matrices at $M_{PS}$, 
according to the above boundary conditions,
we used 422 RGEs (see Appendix 2) to run them to $M_{P}$. 
In this high energy region we treated the theory
described in Section 2 in the following effective way.
To begin with we regarded the non-renormalisable operators
as yielding four effective Yukawa matrices, whose RG evolution 
is described by standard RGEs appropriate to the larger gauge group
\422 . The terms involving the singlet $S$ which give rise to an effective
$\mu$ parameter below the scale $M_{PS}$, were regarded as an
effective $\mu$ parameter above this scale similar to the MSSM.
Finally we allowed extra $D$ and other 
superfields to be present above the scale $M_{PS}$ in order to
keep the one-loop beta functions of the \422 gauge group
equal above this scale, and so 
allow string gauge unification at $M_P$ \cite{king}. 
Since such additional
superfields do not couple to the quark and lepton superfields their
presence will have no effect on the LFV predictions at the one-loop level,
apart from the indirect effect via the gauge couplings. 

At $M_P$ boundary conditions were chosen to reduce the most the number of
independent parameters : $M_i=M_{1/2}$ (common gaugino masses),
$\tilde{m}_i^2=m_{H_{u,d}}^2=m_0^2$ (universal soft masses), 
$\tilde\lambda_i=A \lambda_i$ (proportional soft Yukawa matrices).
With this choice we kept sources of LFV (mass splittings and CKM entries) 
at a minimum. A departure from the latter two conditions would introduce 
right from the start LFV signals therefore the results can be interpreted
as the irreducible minimum amount of LFV arising from this model.

The next obvious step was to run all the above parameters again
down to $M_{SUSY}$, dropping terms from the RGEs involving
right-handed neutrinos and sneutrinos below $M_\nu$. 
At the low energy SUSY scale we were finally able to set 
the superpotential Higgs parameter 
$\mu^{2}$ and soft Higgs mass $\tilde\mu^2$. 
These were the last parameters to be defined because they obey the following
two conditions~:
\begin{equation}
\mu^2 = {m_{H_d}^2-m_{H_u}^2\>\tan^2\beta \over \tan^2\beta-1}-1/2\> m_Z^2
\label{vacummeqs1}
\end{equation}
\begin{equation}
\tilde\mu^2 = 1/2\ (m_{H_u}^2+m_{H_d}^2+2\>\mu^2)\sin 2\beta 
\label{vacummeqs2}
\end{equation}
which depend on the low energy values of $m_{H_u}^2$ and $m_{H_d}^2$ 
until now unknown. The above equations 
simply describe how the Higgs VEVs are related to the 
(classical) renormalised Higgs potential parameters. To see how
they came about we recall that after 
$SU(2)\otimes U(1)_Y\to U(1)_{QED}$ symmetry breaking the neutral
Higgs $H_u^0$, $H_d^0$ acquire VEVs $v_u,v_d$ therefore the Higgs
potential becomes~:
\begin{equation}
{\cal V}(H_u,H_d)\to {\cal V}(v_u,v_d)+
(\hbox{Physical Higgs Interactions})
\end{equation}
\begin{equation}
{\cal V}(v_u,v_d) = (\mu^2+m_{H_u}^2)\>v_u^2 +
                    (\mu^2+m_{H_d}^2)\>v_d^2 -
                    2\>\tilde\mu^2\>v_u v_d +
                    1/8\>(g'^2+g^2)(v_u^2-v_d^2)^2
\end{equation}
In order to recover the traditional interpretation of the VEVs, 
they must satisfy~:
\begin{equation}
{\partial{\cal V}(v_u,v_d) \over \partial v_u} = 0 \qquad
{\partial{\cal V}(v_u,v_d) \over \partial v_d} = 0 \qquad
\end{equation}
Which after simple algebraic manipulation leads to 
(\ref{vacummeqs1},\ref{vacummeqs2}), except for obvious replacement of
$v_u,v_d$ by the more convenient set  $\tan \beta, m_Z$.
From \refeqn{vacummeqs1} we see that $\mu$ is determined up to a sign, 
however we found that for large $\tan \beta$ this 
arbitrariness was not relevant.

The physical origin of LFV in this model is now clear.
The Yukawa coupling matrices are effectively split
into $\lambda_u$, $\lambda_d$, $\lambda_e$, $\lambda_\nu$. These
matrices are not equal but related to each other by different
Clebsh factors. Since 
$\lambda_e \ne \lambda_\nu$ one is introducing a CKM like mixing
matrix on the lepton sector which gets imprinted onto the left-handed
slepton masses due to the RG running between $M_P$ and $M_{\nu}$
(below the scale $M_{\nu}$ the terms involving the right-handed
neutrinos are dropped from the RGEs). 
It is clear that, for example in a basis in which the charged lepton matrix 
$\lambda_e$ is diagonal the neutrino matrix $\lambda_\nu$ will be
non-diagonal, leading to off diagonal contributions to the left-handed
slepton masses of the form
\footnote{There is also an F-term contribution of opposite
sign as discussed in Appendix 5.}
\begin{equation}
\Delta \tilde m^2_L \sim - \frac{\ln (M_P /M_{\nu})}{16 \pi^2}
m_0^2 (\lambda^\hc_\nu \lambda_\nu) +\cdots
\label{softmassrge}
\end{equation}

It is interesting to note that in $SU(5)$ LFV develops
differently. In this model, there is no right-handed neutrino, however,
LFV does arise from the presence of Higgs colour triplets which
mediate tree level leptoquark interactions, which again leads
to off-diagonal slepton masses due to RG running between $M_P$ and $M_{GUT}$.
$SO(10)$ is an example in which both the mentioned LFV
processes are active \cite{strumia,arkani}.


\section{The Processes $\mu \to e+\gamma$ and $\tau \to \mu +\gamma$.}

\subsection{Formalism.}

The effective Lagrangian and branch ratio for the decay 
$\mu\to e+\gamma$ are given by~:
\begin{equation}
{\cal L} = {1 / 2}\>{\bar u}_e(p-q) 
           \left\{ A_R P_R + A_L P_L\right\} 
           \sigma^{\alpha\beta}\>u_\mu(p)\>{\cal F}_{\alpha\beta}
\end{equation}
\begin{equation}
\hbox{BR}(\mu\to e+\gamma)={12\pi^2\over G_F^2 m_\mu^2}
                            \>(\>\vert A_R\vert^2+\vert A_L\vert^2\>)
\end{equation}
\medskip
\noindent
In their most general form, the one-loop amplitudes $A_R=\sum A_{R_i}$,
$A_L=\sum A_{L_i}$ are given by a sum of many terms \cite{hisano} 
most of which of negligible importance. 
For sake of simplicity we consider only the dominant contributions~:
\begin{eqnarray}
A_{R_1} &=& \phantom{-}
            {e \over 16\pi^2} {x \over \sqrt 2 } \>
            (\tilde U_{LL}^{n\hc})_{eA}
            (\tilde U_{LL}^n)_{A\mu} J_{21A} \label{AR1} \\
A_{R_2} &=& -{e \over 16\pi^2} {x \over 2 } \>
            (\tilde U_{LL}^{l\hc})_{eA}
            (\tilde U_{LL}^l)_{A\mu} (H_{32A}+H_{31A}) \\
A_{R_3} &=& -{e \over 16\pi^2} \> \{
           (\tilde U_{LL}^{l\hc})_{eA}
            (\tilde U_{LR}^l)_{A\mu} H_{11A} +
            (\tilde U_{RL}^{l\hc})_{eA}
            (\tilde U_{RR}^l)_{A\mu} H_{11\dot A} \} \\
A_{L_1} &=& \phantom{-} 
            {e \over 16\pi^2} \> x \>
            (\tilde U_{RR}^{l\hc})_{eA}
            (\tilde U_{RR}^l)_{A\mu} H_{31\dot A} \\
A_{L_2} &=& -{e \over 16\pi^2} \> \{
            (\tilde U_{LR}^{l\hc})_{eA}
            (\tilde U_{LL}^l)_{A\mu} H_{11A} + 
            (\tilde U_{RR}^{l\hc})_{eA}
            (\tilde U_{RL}^l)_{A\mu} H_{11\dot A} \label{AL2} \} 
\end{eqnarray}
\medskip
\noindent
here written in a notation, which we now explain.
In all expressions summation over the family index $A=1\ldots 3$ 
is to be understood ($\dot A = A+3$). 
The factor $x = m_\mu / (\cos\beta \ m_W)$ 
and the matrices $\tilde U$ and the other factors occuring in these
expressions are defined in Appendix 3.

We now discuss the phenomenology of equations \refeqn{AR1}-\refeqn{AL2}.
The amplitude $A_L\ll A_R$ because in the 422 model LFVs associated
with $\tilde m_{e^c}^2$ are negligible compared with the ones due to
$\tilde m_L^2$. Thus we will keep our attention on 
$A_{R_1}$, $A_{R_2}$ and $A_{R_3}$ which are associated with the diagrams
in Figure \FigSusyDiag.

\vspace{0.5in}

\vbox{
\hfil
\hbox{
\epsfxsize=4cm
\epsfysize=2.0cm
\epsffile{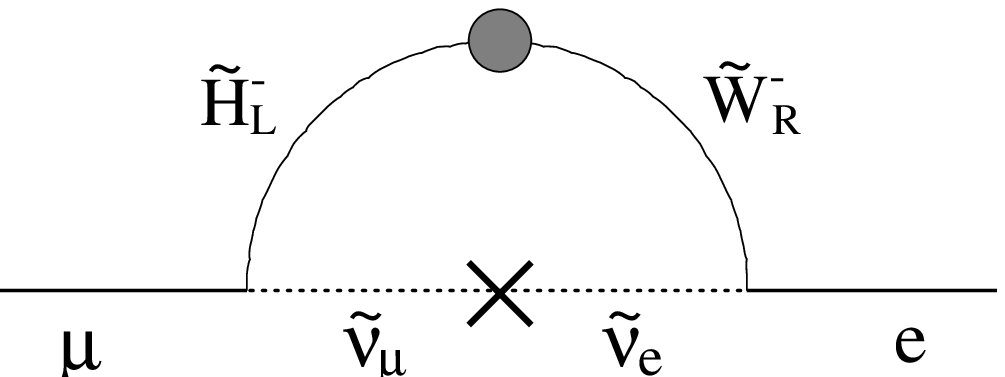} } 
\hfil
\hbox{
\epsfxsize=4cm
\epsfysize=2.0cm
\epsffile{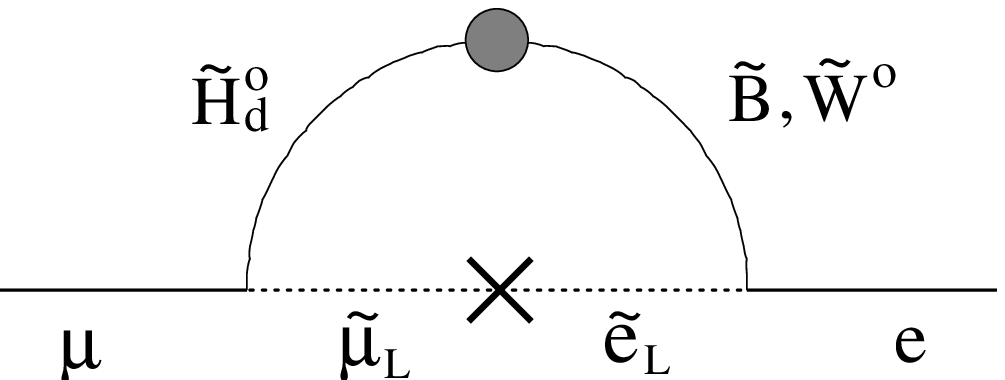} }  
\hfil
\hbox{
\epsfxsize=4cm
\epsfysize=2.0cm
\epsffile{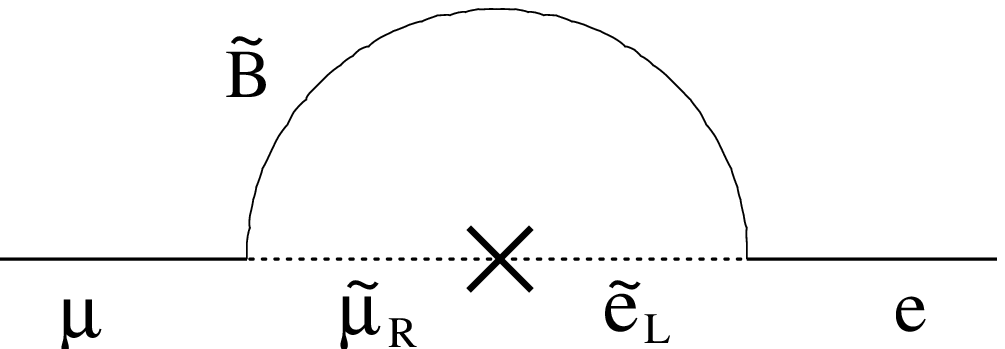}} }
\medskip
\vbox{
{\narrower\narrower\footnotesize\noindent
{\bf Figure \FigSusyDiag.} Dominant supersymmetric diagrams involved in the
decay \muegamma.\par
\bigskip}}

\vspace{0.5in}

In these, a cross over slepton (dotted) line is introduced to remind us of a
$\tilde U \> \tilde U$ product dependence. Similarly,
the blob over chargino-$A_{R_1}$ (neutralino-$A_{R_{2,3}}$) 
line stands for $S^C\> T^C$ ($S^N \> S^N$) dependence.
We stress that, though it is tempting to make a straight analogy with 
the perturbative mass insertion method valid when $\tan \beta$ is small,
such comparison must be taken with great care. For example, 
if $\tan \beta$ is big $A_{R_3}$ will have not only a 
$({\tilde U}_{LL}^{l\hc})_{e\mu}({\tilde U}_{LR}^{l})_{\mu\mu}$
contribution but 
$({\tilde U}_{LL}^{l\hc})_{ee}({\tilde U}_{LR}^{l})_{e\mu}$ 
as well. Nevertheless Figure \FigSusyDiag\ is useful to the extent it 
identifies which supersymmetric states are directly involved
in each diagram.

The neutralino contributions are $A_{R_2}$ which
describes LFV arising from left-handed 
selectrons (\hbox{$\tilde U_{LL}^l=\tilde S_{LL}^l T^{e\hc} \ne 1$}) and
$A_{R_3}$ which is related with mixing of chirality 
($\tilde U_{LR}^l=\tilde S_{LR}^l S^{e\hc}$,
 $\tilde U_{RL}^l=\tilde S_{RL}^l T^{e\hc}$).  
In models with large $\tan\beta$ the slepton mass eigenstates 
$\tilde l_{\tau_L}$, $\tilde l_{\tau_R}$ pick substantial
contributions from supersymmetric states
$\tilde\tau_R$, $\tilde\tau_L$ respectively.
Pertubative expansion of $6\times 6$ matrix $\tilde M^{l2}$ on 
$\tilde M_{LR}^{l2}$, $\tilde M_{RL}^{l2}$ sectors is no longer valid. 
As a consequence full diagonalisation of $\tilde M^{l2}$ renders 
$\tilde S_{LR}^l$ ($\tilde S_{RL}^l$) misaligned with $S^e$ ($T^e$).
Chirality flip is suppressed due to small $\tilde M_{LR}^{l2}$,
$\tilde M_{RL}^{l2}$ but enhanced by diagonal $H_{11A}$ entries. 
These two factors balance each other and make $A_{R_3}\sim A_{R_2}$
(Figure \FigAr).
Lets consider the chargino contribution $A_{R_1}$.
In the large $M_{1/2}$ limit 
$\tilde M_{LL}^{n 2} \sim \tilde  M_{LL}^{l 2}$
and $J\sim H \Rightarrow 
\vert A_{R_1} \vert \sim \vert A_{R_2} \vert \sim \vert A_{R_3} \vert$. 
All flavour violations are due to $\tilde T^L T^{e\hc} \ne 1$. If $M_{1/2}$ is
not too big, $A_{R_1}$ is enhanced because
$\tilde S_{LL}^n$ becomes between $\tilde T^L$ and $T^\nu$ thus
increasing $\tilde S_{LL}^n T^{e\hc}$. Numerically we found $A_R$ to depend
heavily on $A_{R_1}$ over almost all the parameter space studied 
(Figure \FigAr).

\vbox{
\noindent
\hfil
\vbox{
\epsfxsize=8cm
\epsfysize=8cm
\epsffile[160 460 580 720]{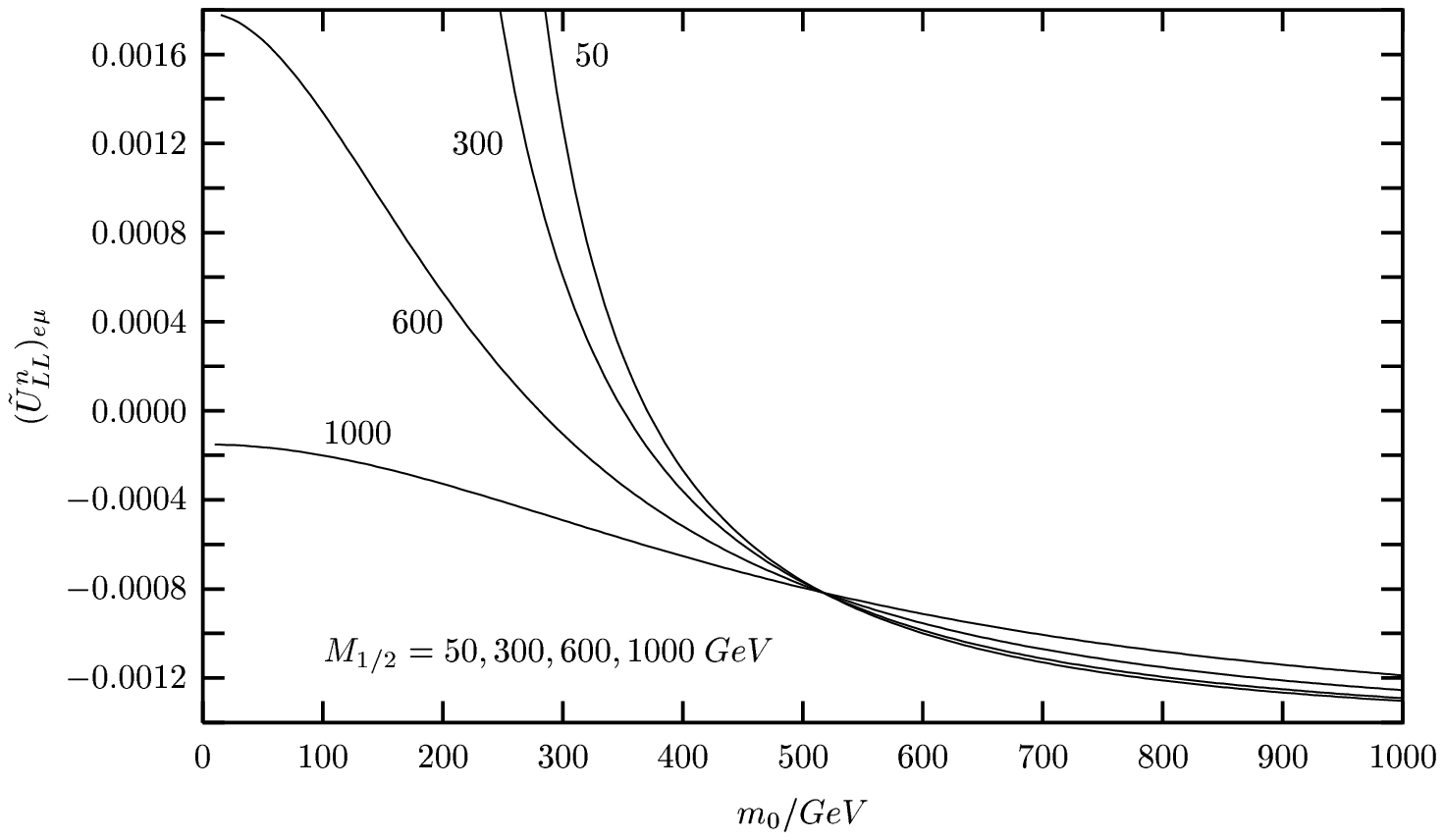}}

{\narrower\narrower\footnotesize\noindent
{\bf Figure \FigUll.} Dependence of $(\tilde U_{LL}^n)_{e\mu}$ with the
universal Planck scalar mass $m_0$ for several values of
gaugino $(M_{1/2})$ mass. The sudden increase in $(\tilde
U_{LL}^n)_{e\mu}$ for $M_{1/2}=50,300$ GeV near $m_0\sim 250$ GeV has little
physical meaning because it relates to the high degeneracy of the
sneutrino spectrum (Figure \FigSneutrinos).\par}}

\vbox{
\noindent
\hfil
\vbox{
\epsfxsize=8cm
\epsfysize=8cm
\epsffile[160 460 580 720]{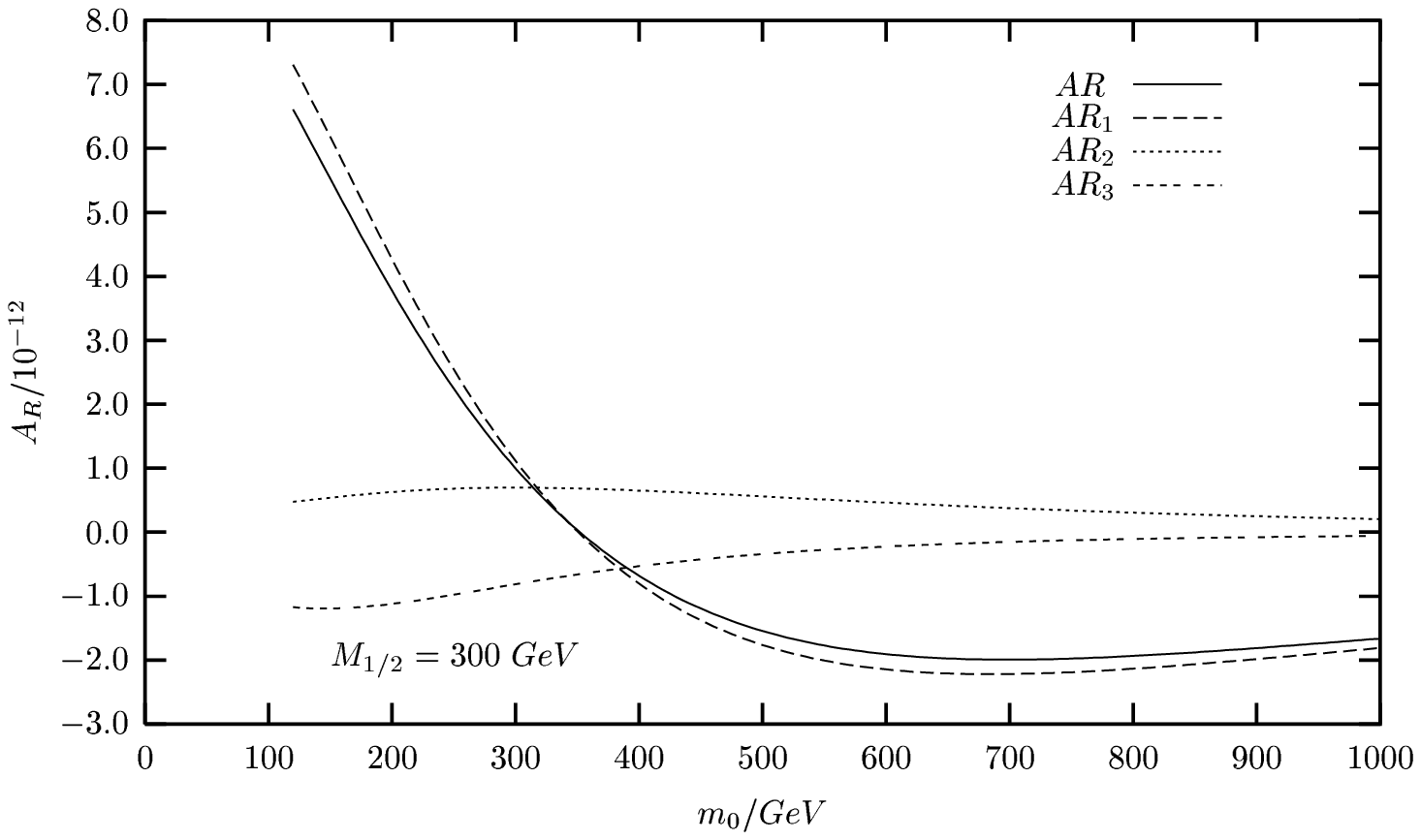}}

{\narrower\narrower\footnotesize\noindent
{\bf Figure \FigAr.} Dependence of dominant amplitudes $A_{R_i}$ with $m_0$.
It is clear that the neutralino contributions $A_{R_{2,3}}$ 
can be neglected over almost all $m_0$ range.\par}
\bigskip}

\vbox{
\noindent
\hfil
\vbox{
\epsfxsize=8cm
\epsfysize=8cm
\epsffile[160 460 580 720]{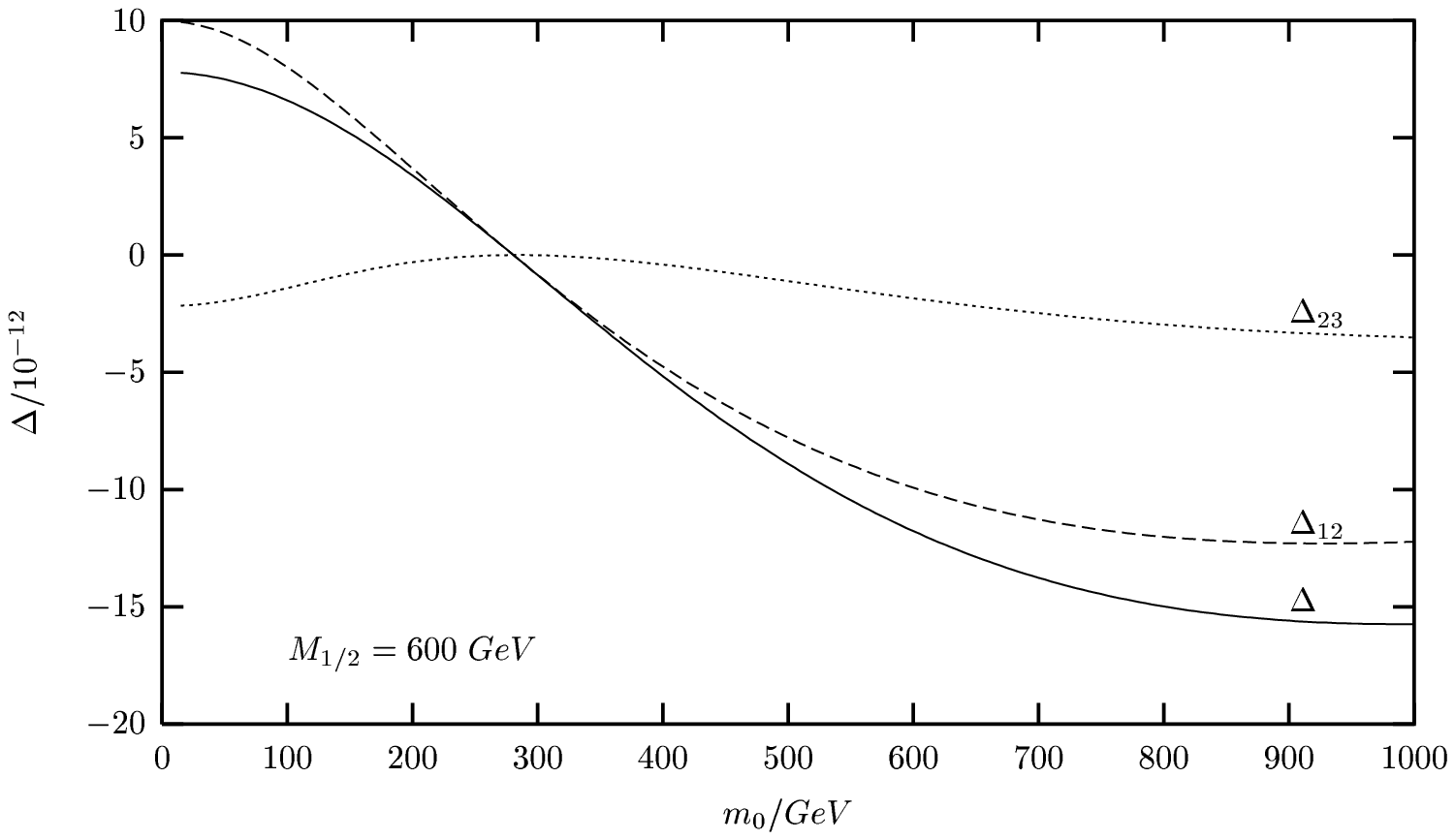}}

{\narrower\narrower\footnotesize\noindent
{\bf Figure \FigDelta.} Lepton flavour violation is driven by
splitting of masses between the first two families as given 
by $\Delta_{12}$.\par}
\bigskip}

\vbox{
\noindent
\hfil
\vbox{
\epsfxsize=8cm
\epsfysize=8cm
\epsffile[160 460 580 720]{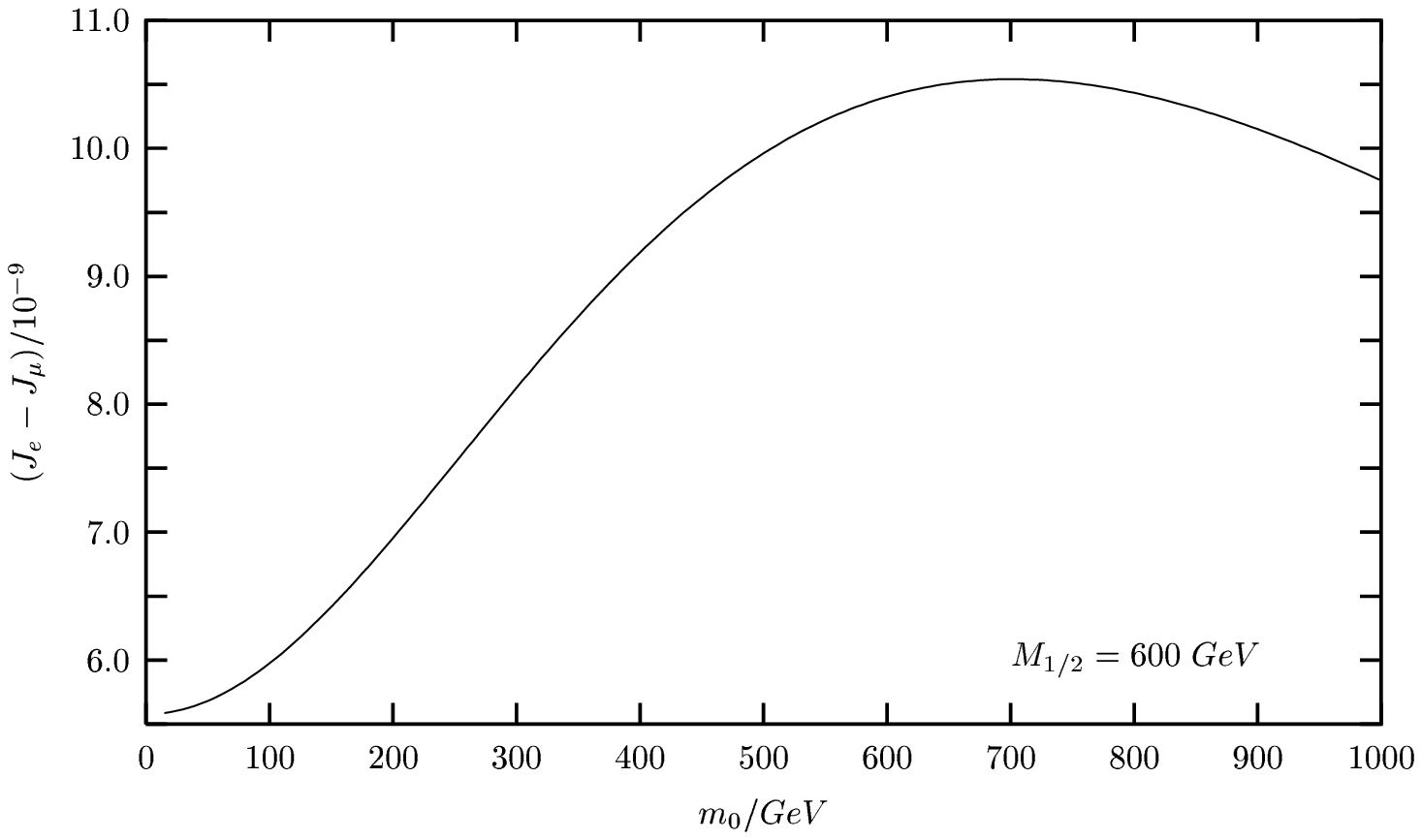}}

{\narrower\narrower\footnotesize\noindent
{\bf Figure \FigJ.} Scaling of $J_e-J_\mu$ function with $m_0$.\par}
\bigskip}

We now focus on the behaviour of $A_{R_1}$ itself. 
Defining $\Delta = \hbox{Const.}\times A_{R_1}=\Delta_{12}+\Delta_{23}$
($J_A = J_{21A}$ and using approximate unitarity of $\tilde U_{LL}^n$) gives~:
\begin{equation}
\Delta_{12} = -(\tilde U_{LL}^{n\hc})_{ee}
               (\tilde U_{LL}^n)_{e\mu} (J_e-J_\mu)
\label{12}
\end{equation}
\begin{equation}
\Delta_{23} = -(\tilde U_{LL}^{n\hc})_{e\tau}
               (\tilde U_{LL}^n)_{\tau\mu} (J_\mu-J_\tau)
\label{23}
\end{equation}
\medskip
\noindent
The above results allow a direct interpretation of LFV in terms of
$e-\mu$, $\mu-\tau$ splittings in Figure 5 and
$J_e - J_\mu $ whose scaling with $m_0$ is shown in
Figure 6. If $\tan\beta$ is small then
$|\Delta_{23}| \gg |\Delta_{12}|$, while as Figure 5 shows
for large \tanb\ over much of parameter space
we find $|\Delta_{12}| \gg |\Delta_{23}|$.
The empirical effect we observe
seems to be related to the large \tanb\ result that
$\lambda_\mu \gg \lambda_{charm}$, which tells us that second family
Yukawa couplings receive an overall enhancement and this is responsible for
the increase in 12 family splittings. It turns out that
the 23 family mixing is also
substantially increased in this model,
but for \muegamma this effect is killed by the small 13 family mixing factor. 
For \taumugamma , on the other hand,
the rate here is controlled exclusively by
23 family mixing and we find a large enhancement in this case.

\subsection{Overview of Results.}

In our numerical results we assumed
$\alpha_s = 0.115$ and $m_{bottom} = 4.25$ \GeV. An increase in
$\alpha_s$ $( m_{bottom})$ leads to smaller slepton masses (bigger
$e-\mu$ splitting) therefore to an enhancement of LFV. 
The parameters we made to vary were $M_{1/2}$, $m_0$, $A_0$ and
$M_\nu$. When not explicitly mentioned the graphs refer to
default values of $A_0 = 0$ \GeV\ and $M_\nu = M_{GUT}$. 
In Figure \FigSleptons\ we plotted the slepton spectrum $\tilde m_l$. 
Due to large $\tan \beta$, for decreasing $m_0$, we verify that the
lightest slepton $\tilde l_{\tau_R}$ is rapidly driven negative,
on the other hand $\tilde l_{\tau_L}$ is pushed upwards and forced to 
be the heaviest sparticle. This phenomena are absent in conventional
low $\tan \beta$ models. 
Figure \FigSneutrinos\ which displays sneutrino masses 
$\tilde m_n$ is interesting because it shows that we do not always
have $\tilde m_{n_e} \sim \tilde m_{n_\mu} > \tilde m_{n_\tau}$.
This relation is inverted for low $m_0$ due to the reverse hierarchies
in the matrices $\tilde m_L$ and $m_\nu$. The neutralino particle
spectrum is shown in Figure \FigNeutralinos.
\footnote{When $m_0$, $M_{1/2} > m_Z$, the heaviest chargino is
approximately degenerate with the two heaviest neutralinos (which
scale with $\mu$, the Higgs mixing parameter), while the
lightest chargino has the same mass as the second lightest neutralino 
(both scale with $M_2$). Finally the lightest neutralino scales with $M_1$.
These three sets are in close correspondence with 
the scaling of Higgsinos, Winos and Bino.}

\vbox{
\hfil
\vbox{
\epsfxsize=8cm
\epsfysize=8cm
\epsffile[160 460 580 720]{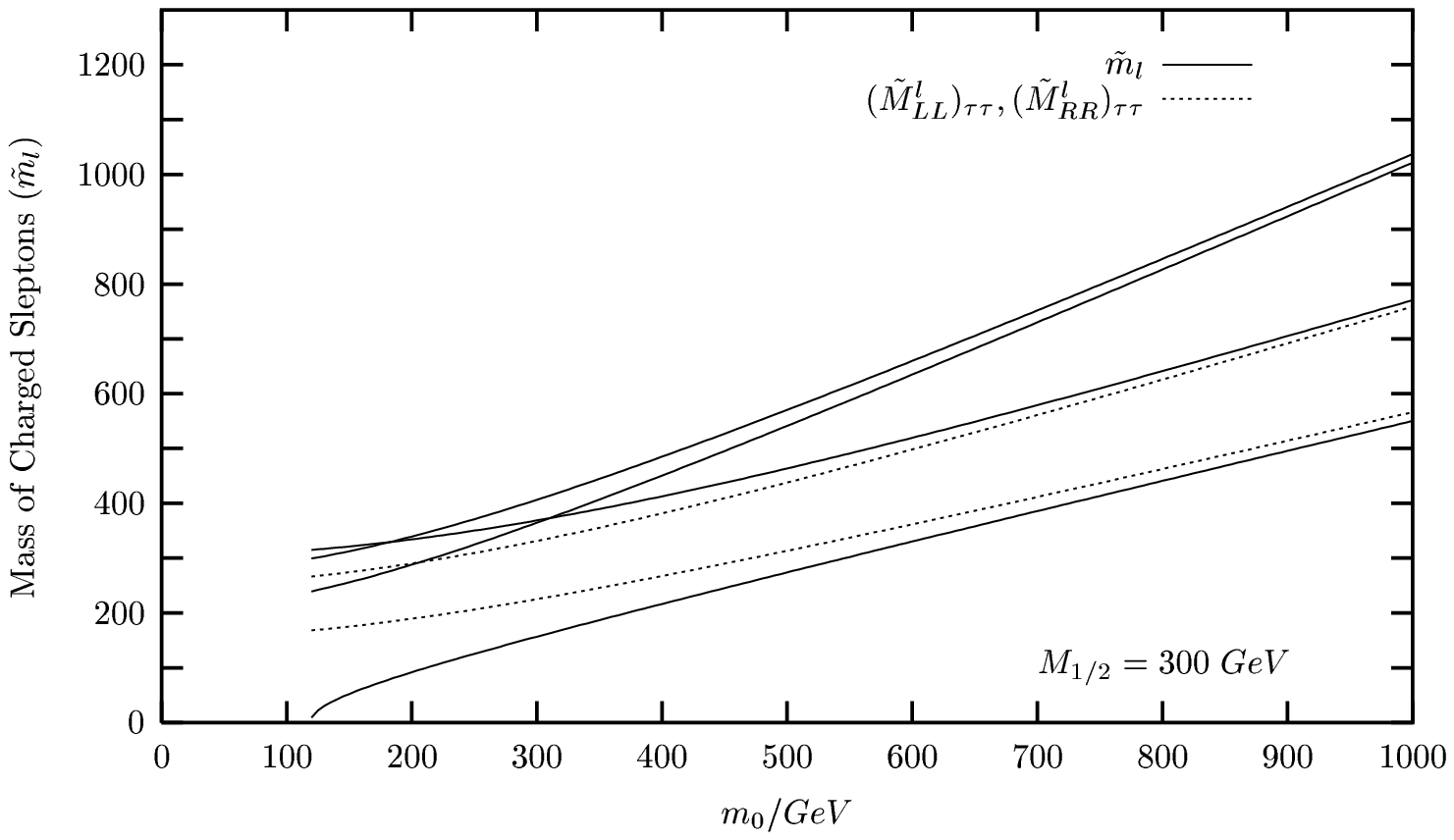}}

{\narrower\narrower\footnotesize\noindent
{\bf Figure \FigSleptons.} Spectrum of charged sleptons $\tilde l$ for a
range of $m_0$ and a selected value of gaugino mass $M_{1/2}$. 
Mixing between left and right staus is shown as a deviation
of the solid line from the dotted one.\par
\bigskip}}

\vbox{
\hfil
\vbox{
\epsfxsize=8cm
\epsfysize=8cm
\epsffile[160 460 580 720]{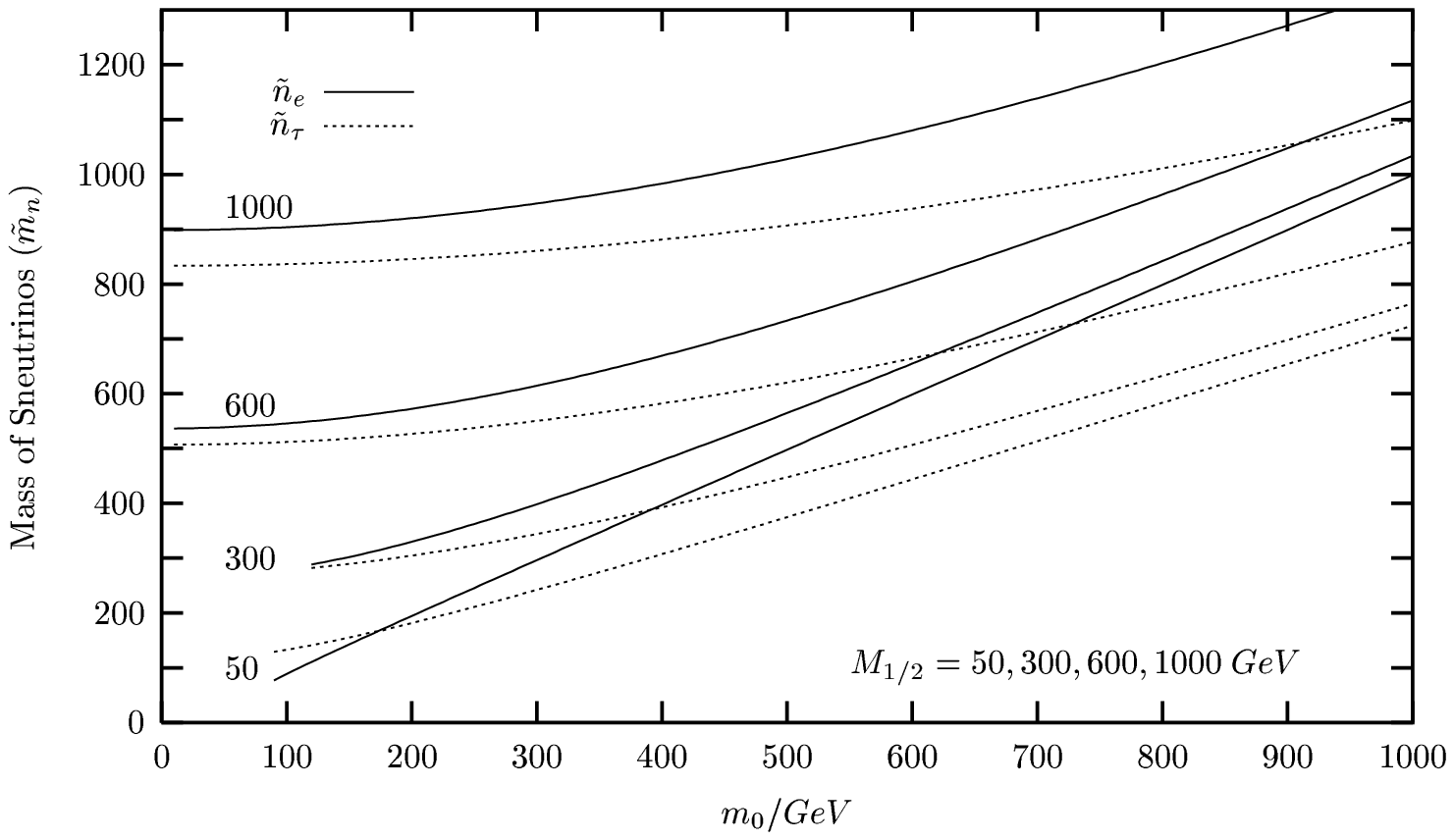}}

{\narrower\narrower\footnotesize\noindent
{\bf Figure \FigSneutrinos.}
Spectrum of sneutrinos $\tilde n_e$ and $\tilde n_\tau$ as
function of $m_0$ for several values of $M_{1/2}$.\par
\bigskip}}

\vbox{
\hfil
\vbox{
\epsfxsize=8cm
\epsfysize=8cm
\epsffile[160 460 580 720]{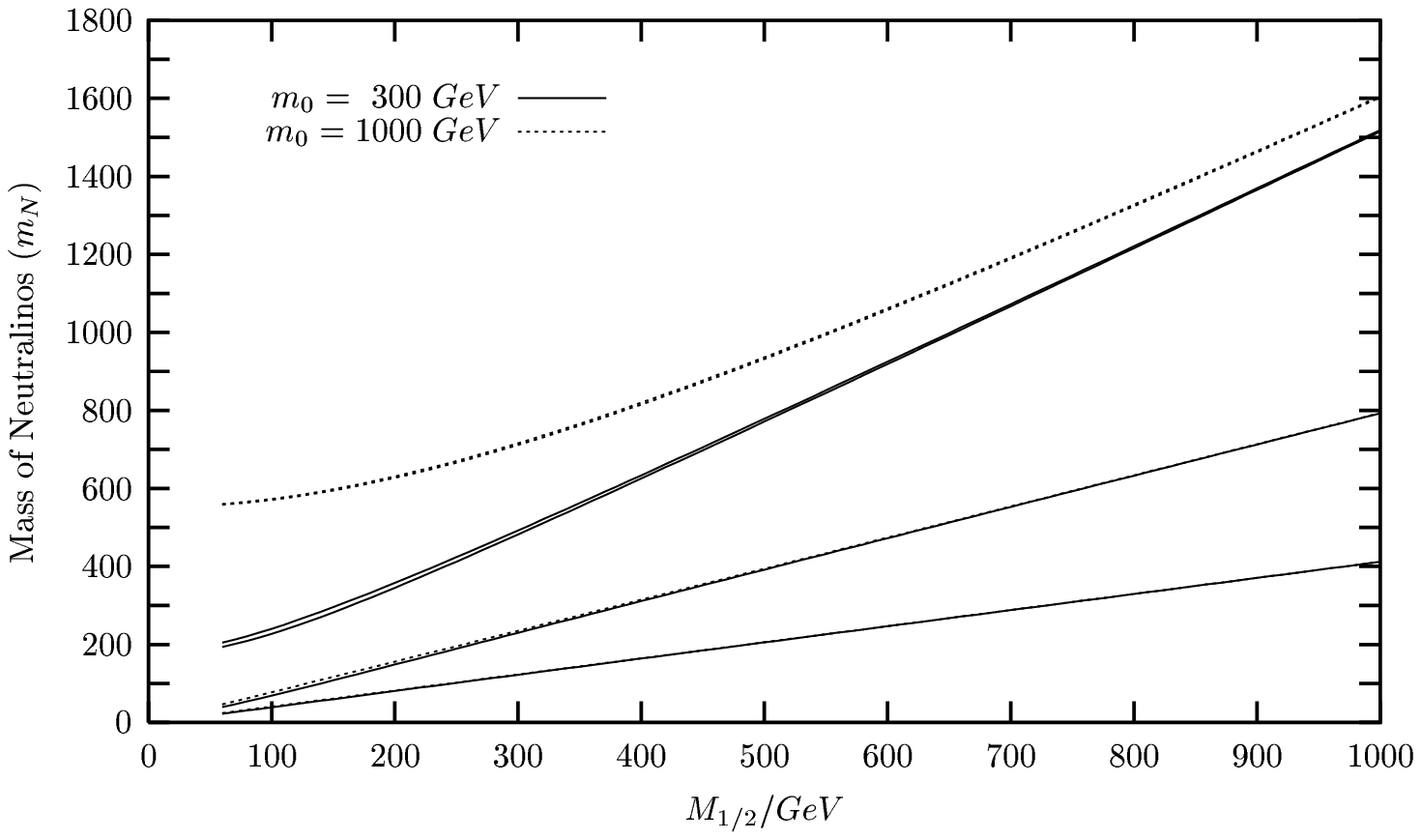}}

{\narrower\narrower\footnotesize\noindent
{\bf Figure \FigNeutralinos.} 
Spectrum of neutralinos as function of $M_{1/2}$.\par
\bigskip}}

\vbox{
\hfil
\vbox{
\epsfxsize=8cm
\epsfysize=8cm
\epsffile[160 460 580 720]{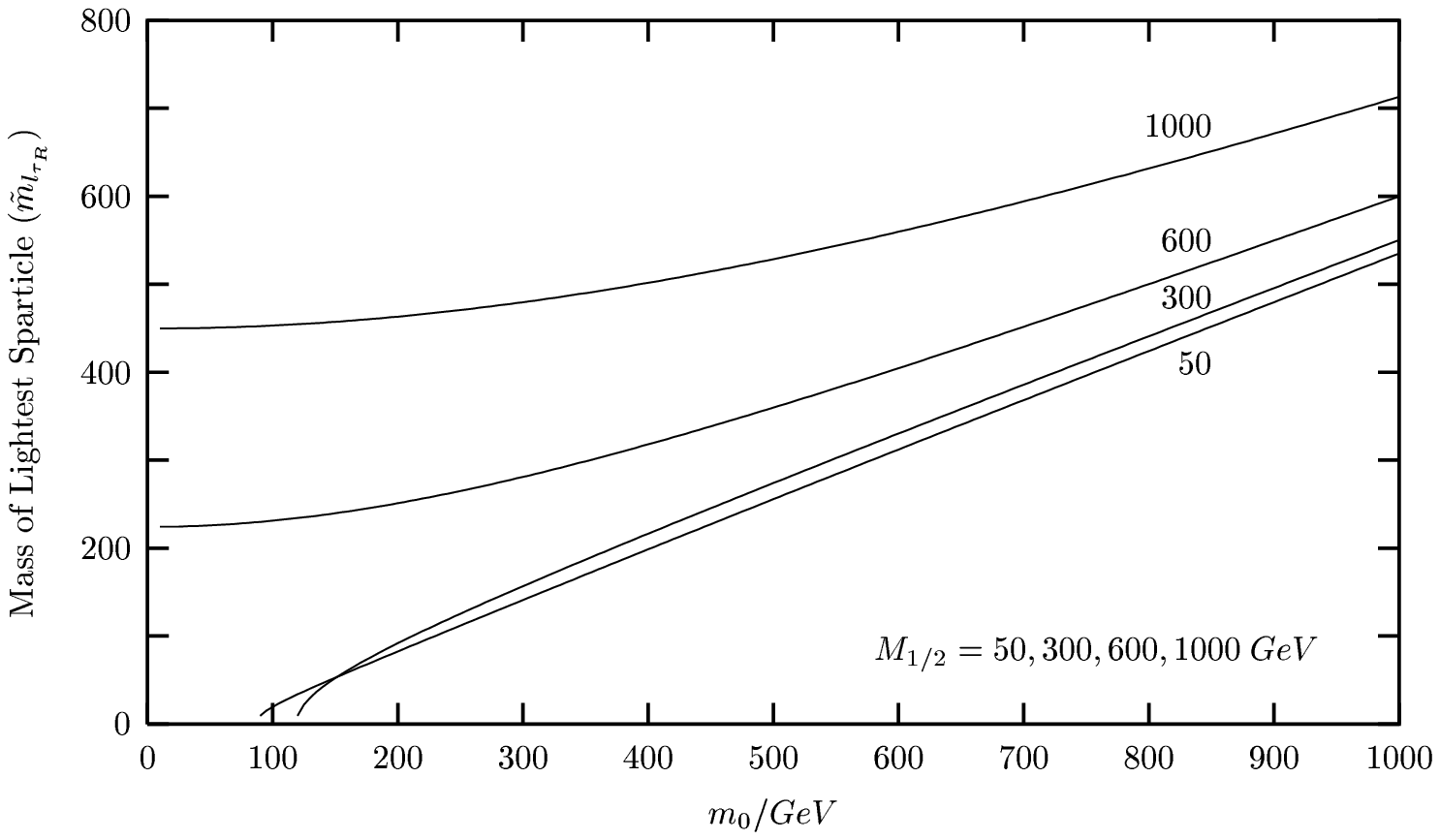}}

{\narrower\narrower\footnotesize\noindent
{\bf Figure \FigLighestScalar.} 
Mass of lightest scalar sparticle, namely the right-handed stau
($\tilde l_{\tau_R}$).\par
\bigskip}}

We now consider the branch ratio $\mu\to e+\gamma$. 
In Figure \FigBRmuegammaMain\ we 
plot its dependence with $m_0$ for selected values of
$M_{1/2}$ ( and just two extreme values of $M_\nu$ ). Discontinued
lines signal regions where $\tilde m_{l_{\tau_R}} \to 0$.
Generally BR($\mu\to e+\gamma$) decreases for increasing $M_{1/2}$
and/or $m_0$ because we are getting heavier sparticles on that
limit. However for the lower $M_{1/2}$ values
the behaviour is clearly not smooth, with a resonant inverted
spike for particular $m_0$ values. This behaviour is due
to the off-diagonal (flavour-violating) elements of the
left-handed sneutrino mass matrix cancelling to zero at particular
points in parameter space, and is discussed in more detail in Appendix 5.
For $M_{1/2} \le 800$ \GeV\ one can split the graph into two regions
according to $m_0$ being bigger or smaller than $\bar m_0$ - the
value of $m_0$
at which $A_{R_1} \sim 0$ ( $\bar m_0$ itself decreases for increasing
$M_{1/2}$ ). The present experimental bound 
BR$(\muegamma) < 4.9\times 10^{-11}$ already excludes regions
of low $( M_{1/2}, m_0 ) < ( 50, 150)$ \GeV. The fact that 
$A_{R_1}$ can be $\sim 0$ makes this model hard to be excluded
as a  possible GUT candidate because even if the experimental
limit on the branch ratio value gets as low as $10^{-16}$ 
one can still argue that we have a very predictive model
\footnote{This is not unique to 422, SU(5) also shares this kind
of behaviour \cite{strumia}, though for a different reason.}.

Figure \FigBRmuegammaMnu\ gives the branch ratio dependence on $M_\nu$,
the variations being due to the presence (absence) of 
right-handed sneutrinos in \MSSM+N (MSSM) above (below) $M_\nu$.
The variations with $A_0$ are plotted in Figure \FigBRmuegammaAo. 
When $\tan \beta$ is big the sparticles mass matrix is explicit independent of
$\tilde\lambda_{e,\nu}$. Since RGEs care only about 
$\tilde\lambda_{e,\nu}^2$ we see that the results must be 
approximately invariant under $A_0 \to -A_0$. For that reason we have
only considered $A_0 > 0$. If $M_{1/2} > 400$ \GeV\
the branch ratio changes only slightly with $A_0$. 
For $M_{1/2}= 300, 400$ \GeV\ increasing $A_0$ drives $A_{R_1}$
to flip its sign, which actually happens for the former case and
doesn't for the latter due to the vanishing of $\tilde l_{\tau_R}$ mass.

In Figure \FigBRtaumugammaMain\ we show the BR(\taumugamma) which is
experimentally constrained to be less than $4.2\times 10^{-6}$. 
This limit demands, for light $M_{1/2} < 50$ \GeV\ 
even heavier SUSY particles than the ones
imposed by \muegamma. For $M_{1/2}$ around/less $50$ \GeV\ we find
$m_0 > 220$ \GeV. On the other hand, 
in the heavier $M_{1/2} > 500$ \GeV\ extreme, a severe
improvement of experimental accuracy (combined with continued
undetection of SUSY sparticles) leads to an upper bound for $m_0$
around 350 \GeV.

\vbox{
\hfil
\vbox{
\epsfxsize=8cm
\epsfysize=8cm
\epsffile[160 460 580 720]{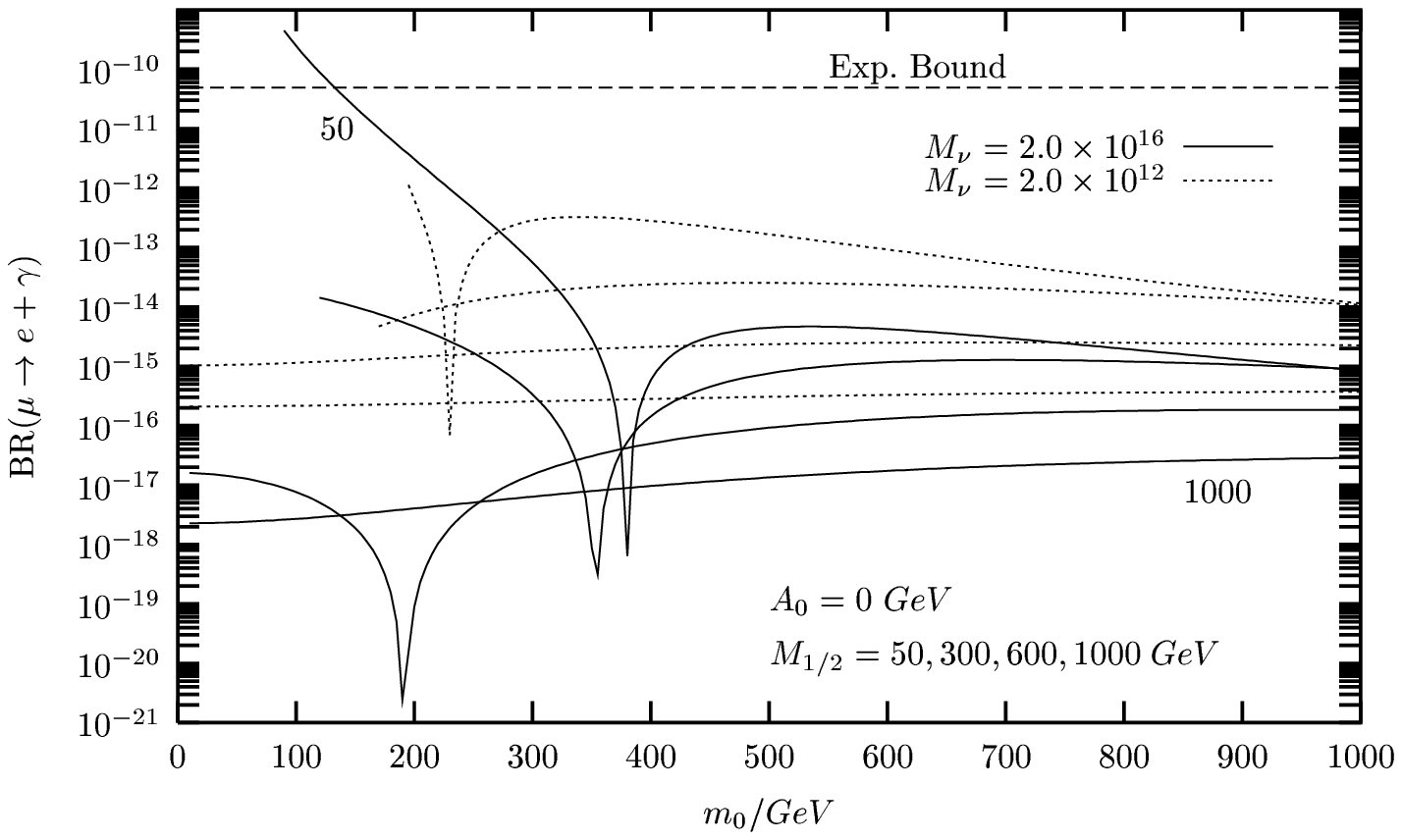}}

{\narrower\narrower\footnotesize\noindent
{\bf Figure \FigBRmuegammaMain.} 
Branch ratio for the decay \muegamma as function of
$m_0$ for several values of $M_{1/2}$ and two $M_\nu$ scales.\par
\bigskip}}

\vbox{
\hfil
\vbox{
\epsfxsize=8cm
\epsfysize=8cm
\epsffile[160 460 580 720]{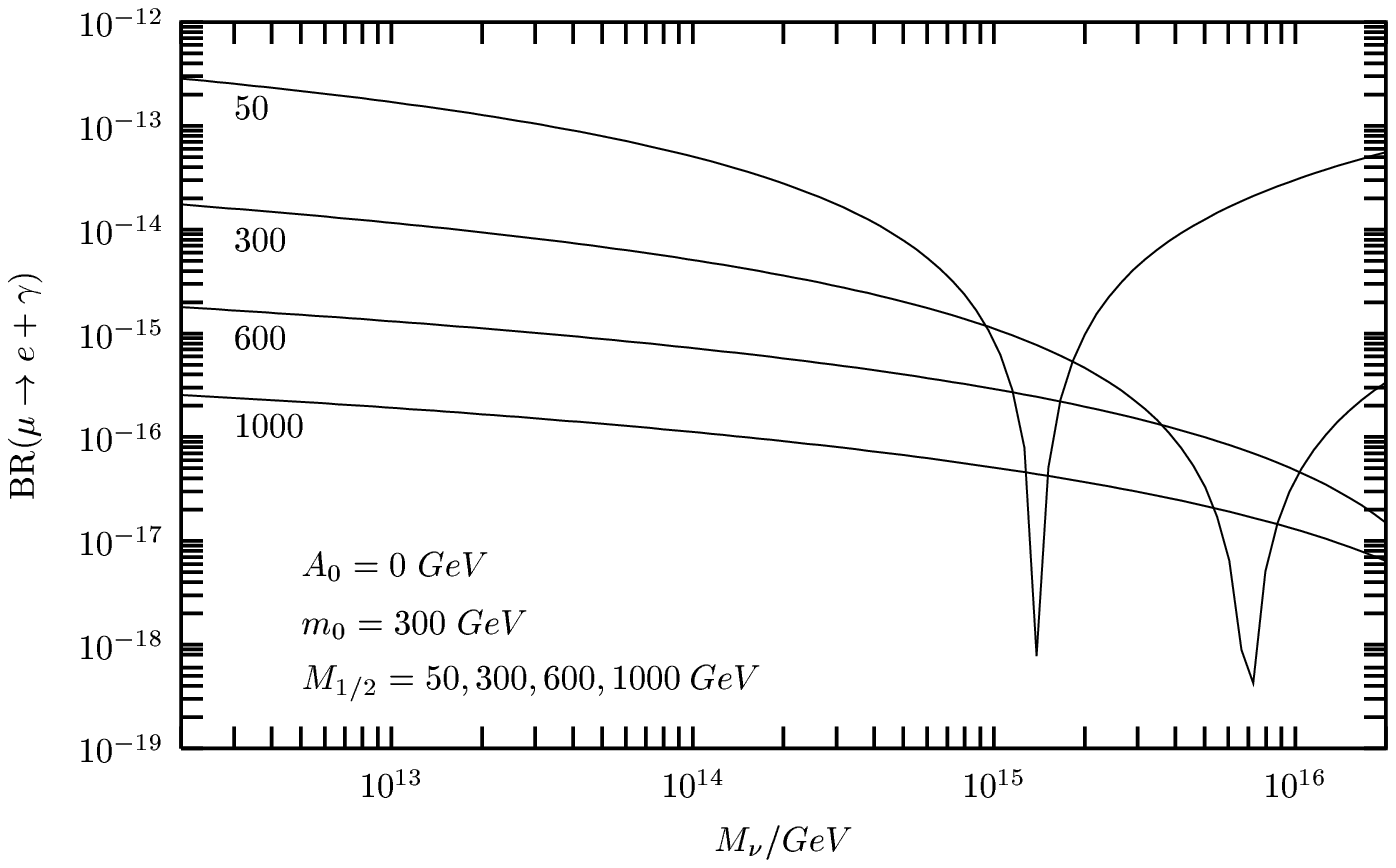}}

{\narrower\narrower\footnotesize\noindent
{\bf Figure \FigBRmuegammaMnu.} Branch ratio for \muegamma versus $M_\nu$.\par
\bigskip}}

\vbox{
\hfil
\vbox{
\epsfxsize=8cm
\epsfysize=8cm
\epsffile[160 460 580 720]{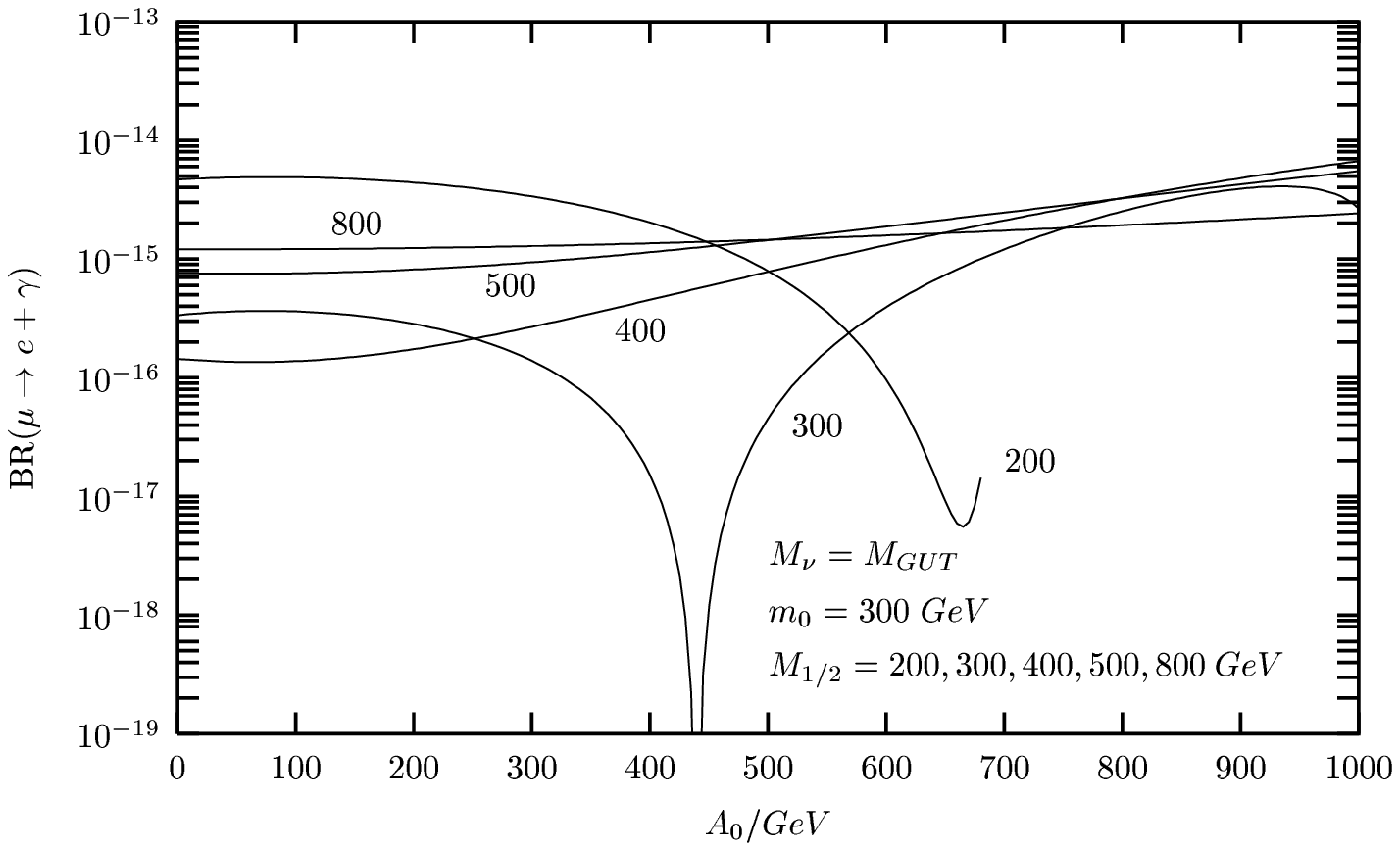}}

{\narrower\narrower\footnotesize\noindent
{\bf Figure \FigBRmuegammaAo.} 
Branch ratio of \muegamma as function of $A_0$.\par
\bigskip}}

\vbox{
\hfil
\vbox{
\epsfxsize=8cm
\epsfysize=8cm
\epsffile[160 460 580 720]{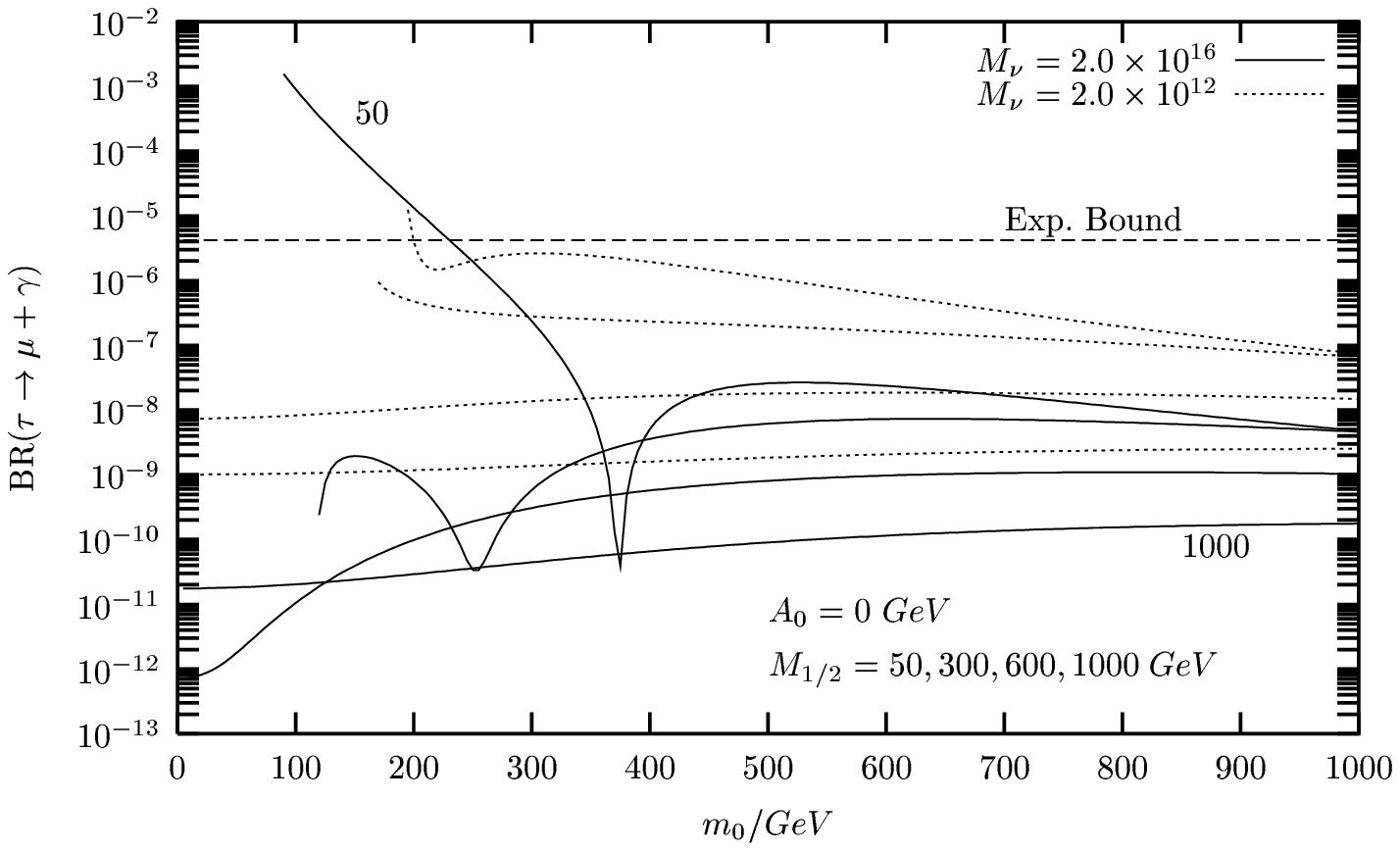}}

{\narrower\narrower\footnotesize\noindent
{\bf Figure \FigBRtaumugammaMain.} 
Branch Ratio of \taumugamma for a range of $m_0$ and
several values of $M_{1/2}$.\par
\bigskip}}

\subsection{Results Near the Experimental Limits.}

It is clear from the results so far that the interesting region
of parameter space from the point of view of the LFV processes
corresponds to relatively low values of soft SUSY breaking
parameters, say $m_0<500 $ GeV and $M_{1/2}<200$ GeV.
In this subsection we shall concentrate on this region, and examine
the relationship between LFV processes and direct experimental
bounds on the particle mass limits coming from LEP, for example.

In Figure \FigNewBRmuegammmaMain\ we show the
branch ratio for \muegamma for a range of $m_0 < 500$ GeV and several
values of $M_{1/2} < 200$ GeV, taking two extreme values of right-handed
neutrino mass.
As the experimental bound improves it is clear how increasingly larger 
regions of the $m_0-M_{1/2}$ plane in this model may be excluded, with the
low values of right-handed neutrino mass (well motivated from neutrino
physics) providing the larger rates closer to the experimental limit.

Figure \FigNewBRtaumugammmaMain\ shows the predicted 
branch ratio for \taumugamma . The well-motivated 
$M_\nu =2\times 10^{12}$ GeV curves are quite close to the
experimental limit, which if increased by an order of magnitude
could provide a decisive test of this model.

Figure \FigNewNewBRtaumugammma\ shows the dependence of \taumugamma on
the right-hand neutrino scale-$M_\nu$. One can see that, as $M_\nu$
decreases, the right-handed neutrino decouples at a lower energy,
therefore allowing additional LFV to be generated through the RGE running,
which leads to an enhancement in the branch ratio. This
effect becomes so strong at low $M_\nu$ that it
overcomes the other competing source of LFV (see Appendix~5).
As a consequence, the inverted peak becomes less pronounced and broader
while at the same time steadily moves to lower $m_0$ until it becomes 
indistinguishable. 

\vbox{
\noindent
\hfil
\vbox{
\epsfxsize=8cm
\epsfysize=8cm
\epsffile[160 460 580 720]{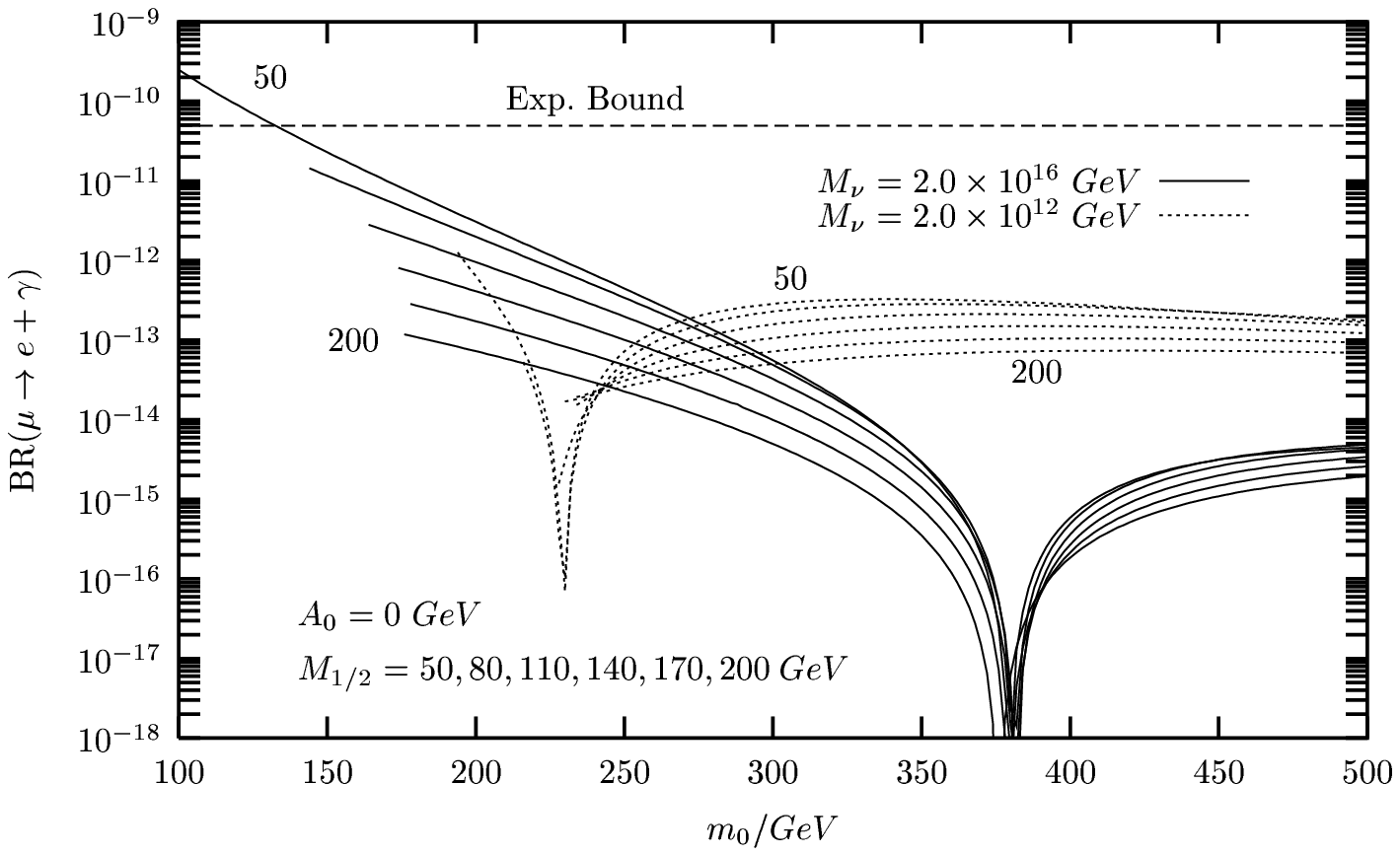}}

{\narrower\narrower\footnotesize\noindent
{\bf Figure \FigNewBRmuegammmaMain.}
Branch ratio for \muegamma for a range of $m_0 < 500$ GeV and several
values of $M_{1/2} < 200$ GeV. Two extreme values of $M_\nu$ are
displayed : solid lines correspond to $M_\nu=M_{GUT}$, while
dotted lines to $M_\nu=2\times 10^{12}$ GeV.\par}
\bigskip}

\vbox{
\noindent
\hfil
\vbox{
\epsfxsize=8cm
\epsfysize=8cm
\epsffile[160 460 580 720]{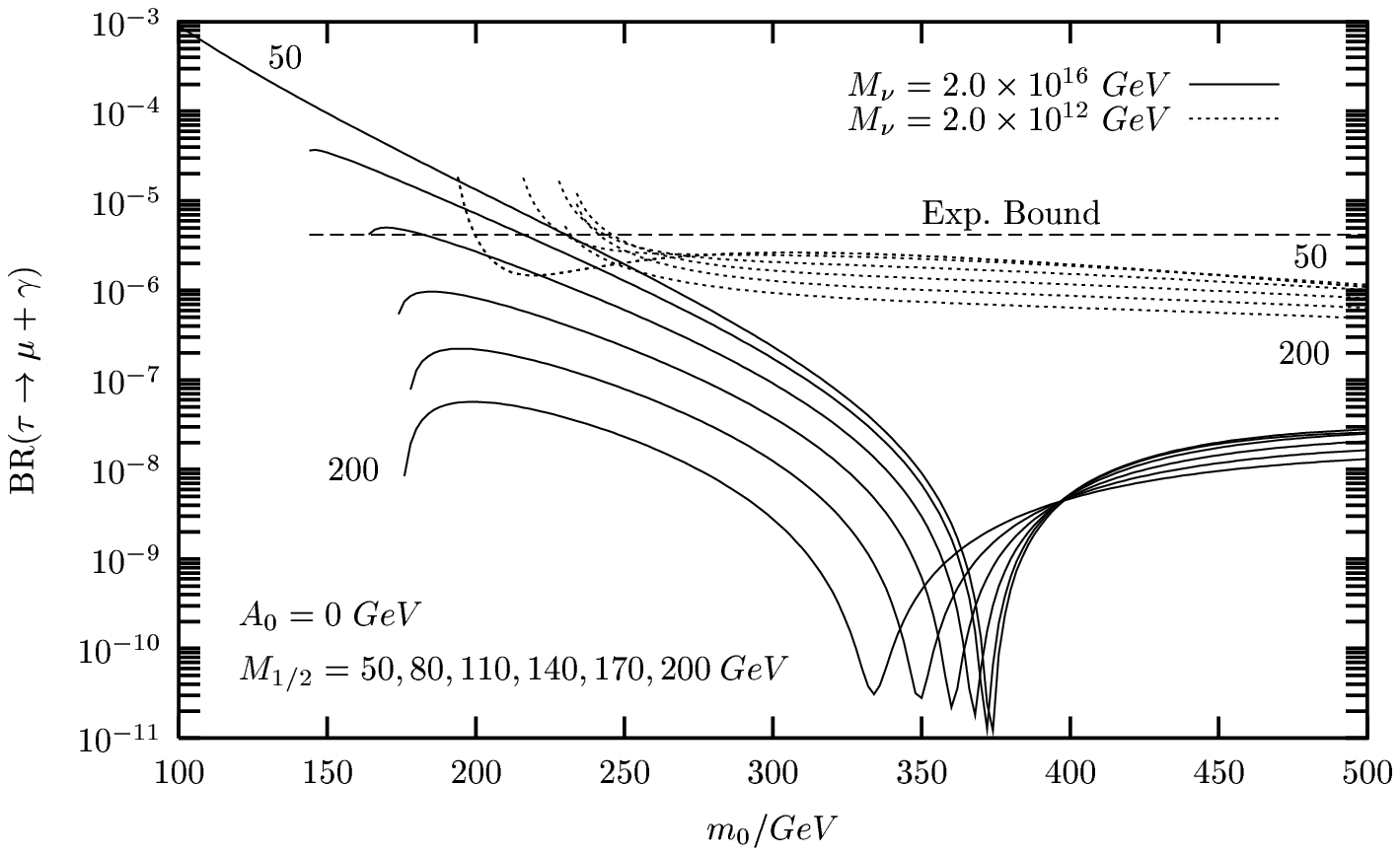}}

{\narrower\narrower\footnotesize\noindent
{\bf Figure \FigNewBRtaumugammmaMain.}
Branch ratio for \taumugamma for a range of $m_0 < 500$ GeV and several
values of $M_{1/2} < 200$ GeV. Two extremes values of $M_\nu$ are
displayed : solid lines correspond to $M_\nu=M_{GUT}$, while
dotted lines to $M_\nu=2\times 10^{12}$ GeV. \par}
\bigskip}

\vbox{
\noindent
\hfil
\vbox{
\epsfxsize=8cm
\epsfysize=8cm
\epsffile[160 460 580 720]{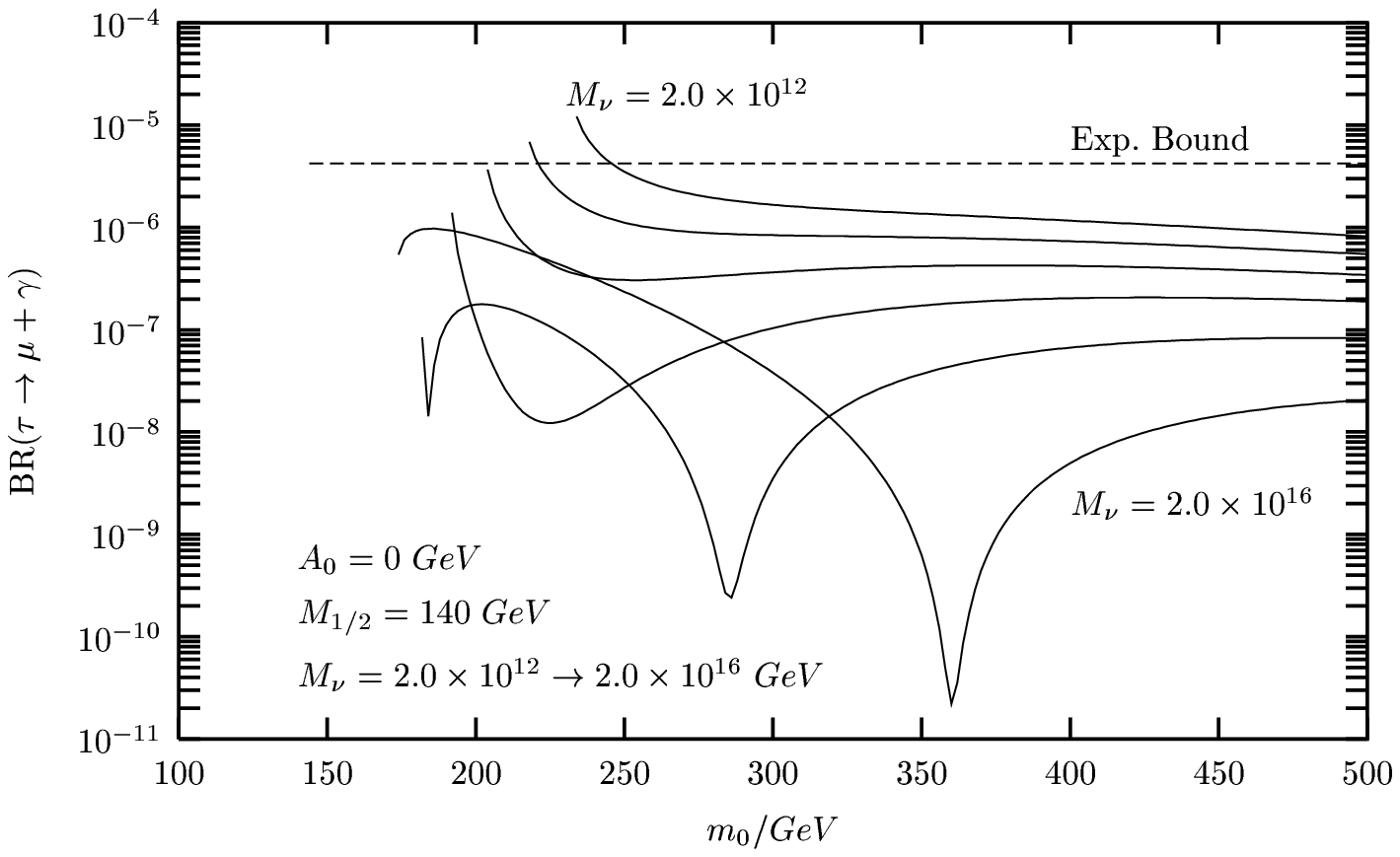}}

{\narrower\narrower\footnotesize\noindent
{\bf Figure \FigNewNewBRtaumugammma.}
Branch ratio for \taumugamma for a range of $m_0 < 500$ GeV 
and $M_{1/2} = 140$ GeV. The six curves plotted correspond to
equally log-scaled intervals of $M_\nu$ in the range 
$2\times 10^{12}$ to $2\times 10^{16}$. \par}
\bigskip}

In this model the spectrum in completely determined by the
values of the input parameters, in particular $m_0$ and $M_{1/2}$,
with very little sensitivity to $M_{\nu}$ for example.
It is clearly of interest to compare the direct experimental
limits which may be placed of these parameters, from LEP for example,
to the indirect limits coming from the LFV processes we have considered.
Therefore we present a series of plots which give a detailed exposition
of the sparticle spectrum in the low mass
region where experiments are sensitive
to LFV processes. 

We begin in Figure \FigNewSleptons\ by showing the spectrum of charged
sleptons for a fixed low value of $M_{1/2}=140$ GeV
corresponding to charginos in the unexplored LEP2
range 95-105 GeV (as we shall see shortly).
The plot shows that the lightest charged slepton mass ranges from
75-250 GeV over the region of $m_0=215-500$ GeV allowed in
a scenario in which the LFV bound for the $\tau$ decay has improved 
to $4.2\times 10^{-7}$ \GeV.
Figure \FigNewLightestScalar\ shows the very weak dependence of the
lightest slepton mass on $M_{1/2}$.
The corresponding sneutrino masses in Figure \FigNewSneutrinos\ 
have a similar mass dependence but are somewhat
heavier.

\vbox{
\noindent
\hfil
\vbox{
\epsfxsize=8cm
\epsfysize=8cm
\epsffile[160 460 580 720]{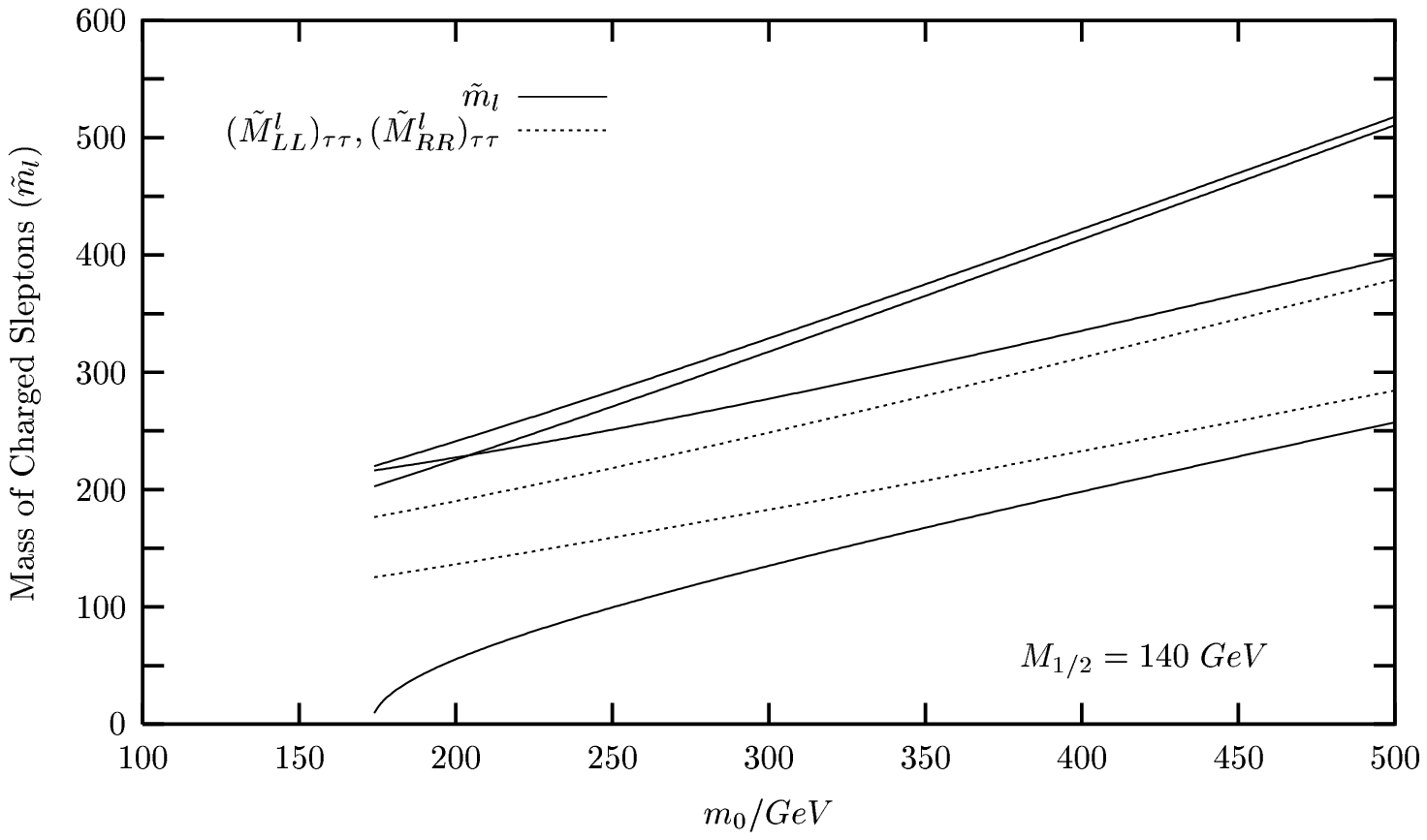}}

{\narrower\narrower\footnotesize\noindent
{\bf Figure \FigNewSleptons.}
Spectrum of charged sleptons ($\tilde l$) for a range of $m_0$ and a fixed
low value of $M_{1/2}=140$ GeV. Mixing between left and right staus is
shown as a deviation of the solid from the dotted one
($A_0=0$ GeV, $M_\nu=2.0\times 10^{16}$ GeV).\par}
\bigskip}

\vbox{
\noindent
\hfil
\vbox{
\epsfxsize=8cm
\epsfysize=8cm
\epsffile[160 460 580 720]{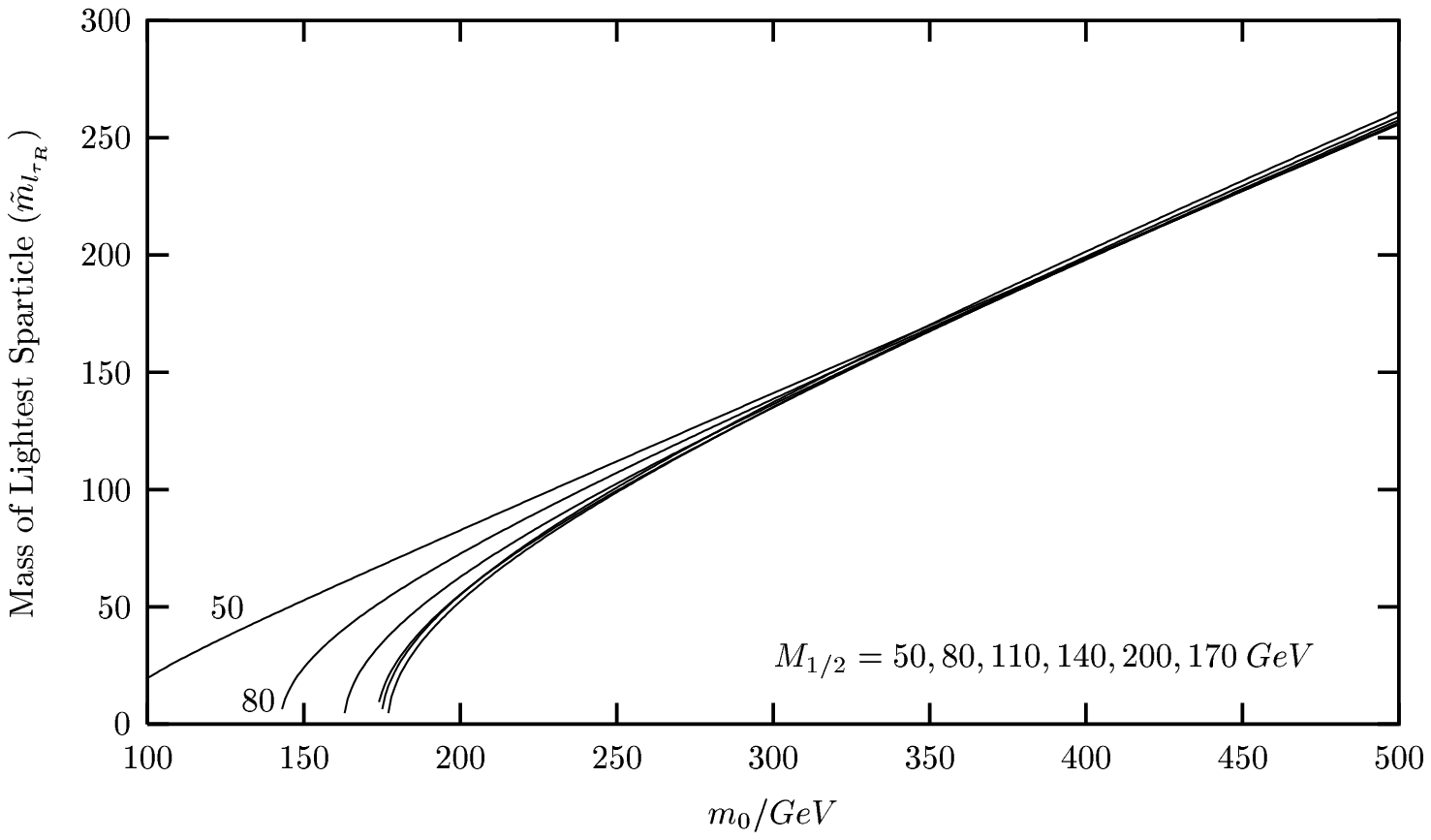}}

{\narrower\narrower\footnotesize\noindent
{\bf Figure \FigNewLightestScalar.}
Mass of lightest sparticle, namely the right-handed stau ($\tilde l_{\tau_R}$),
for a range of $m_0 < 500$ GeV and several values of $M_{1/2}$
(Note that the ordering of the $M_{1/2}$ lines which are plotted, from
left to right, are in correspondence with the order shown on the label in
the graph).\par}
\bigskip}

\vbox{
\noindent
\hfil
\vbox{
\epsfxsize=8cm
\epsfysize=8cm
\epsffile[160 460 580 720]{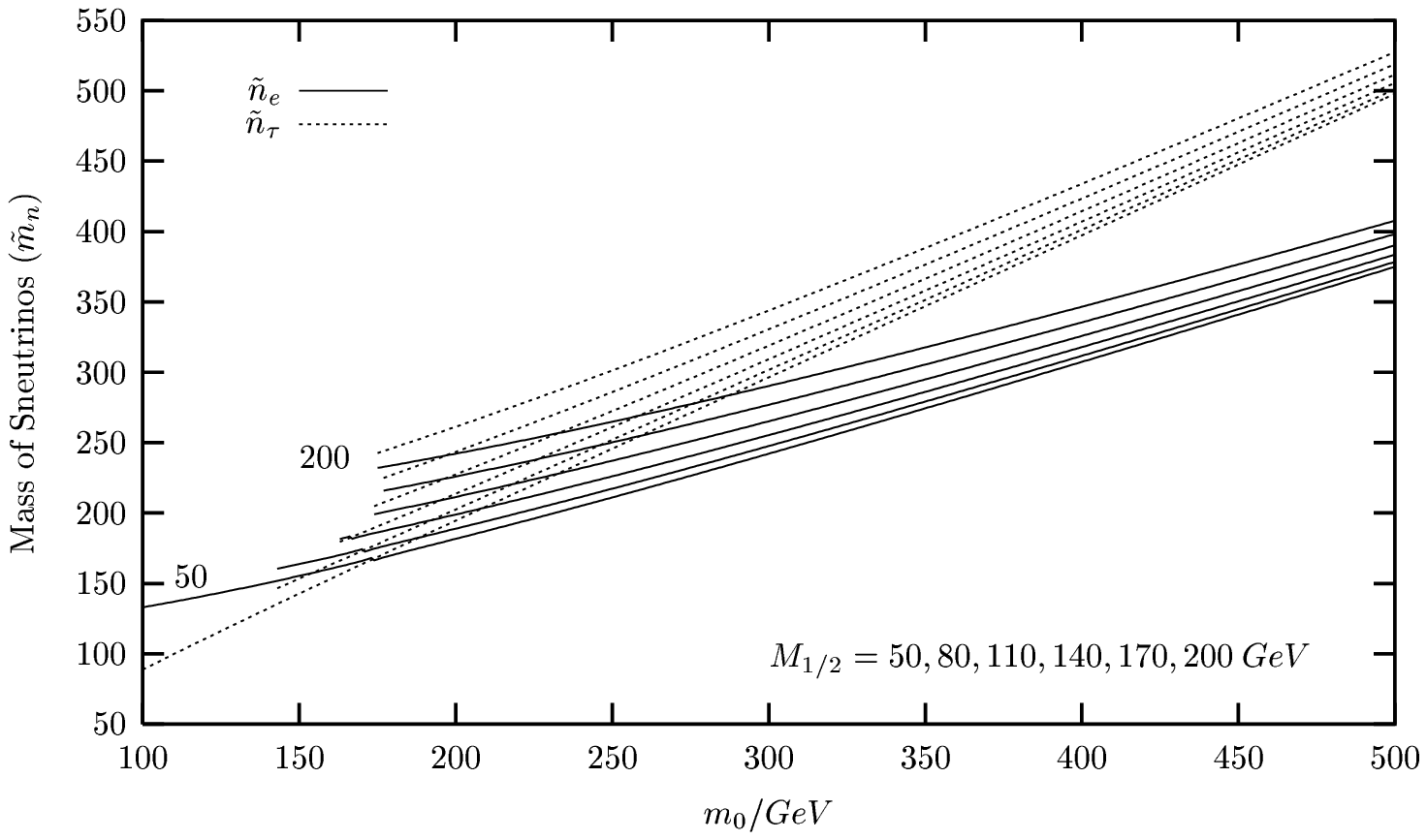}}

{\narrower\narrower\footnotesize\noindent
{\bf Figure \FigNewSneutrinos.}
Spectrum of sneutrinos $\tilde n_e$ and $\tilde n_\tau$ for a range of
$m_0 < 500$ GeV and several values of $M_{1/2}$
($A_0=0$ GeV, $M_\nu=2.0\times 10^{16}$ GeV).\par}
\bigskip}

\vbox{
\noindent
\hfil
\vbox{
\epsfxsize=8cm
\epsfysize=8cm
\epsffile[160 460 580 720]{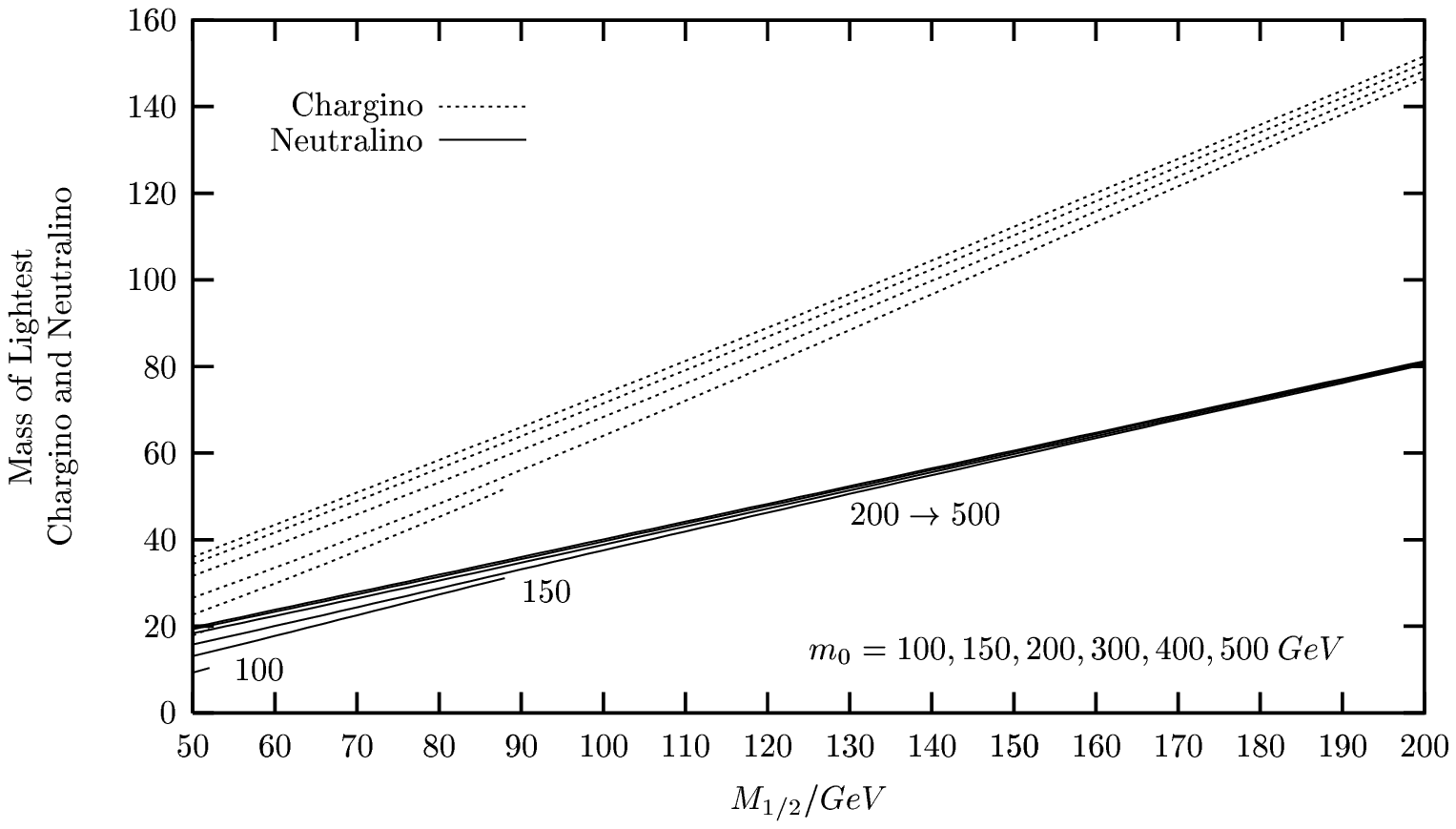}}

{\narrower\narrower\footnotesize\noindent
{\bf Figure \FigNewLightestCharginoNeutralino.}
Spectrum of lightest chargino and neutralino for a range of $M_{1/2}$
and selected values of $m_0 < 500$ GeV 
($A_0=0$ GeV, $M_\nu=2.0\times 10^{16}$ GeV).\par}
\bigskip}

The strongest constraint on $M_{1/2}$ comes from the lightest charginos
and neutralinos in Figure \FigNewLightestCharginoNeutralino.
The full spectrum of charginos and neutralinos, for
a fixed value of $m_0=300$ GeV, and varying $M_{1/2}$, is shown in
Figure \FigNewCharginoNeutralinoB. 

The current published LEP2 limit on chargino masses is around 85 GeV
\cite{ALEPH, DELPHI, L3, OPAL}. This bound does not include 
analysis of the most recent runs, which will increase
this limit to about 91 GeV. A chargino mass limit 
of 91 GeV would correspond to $M_{1/2} > 125$ GeV for $m_0$ in the
intermediate $200-500$ GeV range. The experimental limit on
\taumugamma would need to be increased by one or
two orders of magnitude in order to be competitive with these direct
limits.

\vbox{
\noindent
\hfil
\vbox{
\epsfxsize=8cm
\epsfysize=8cm
\epsffile[160 460 580 720]{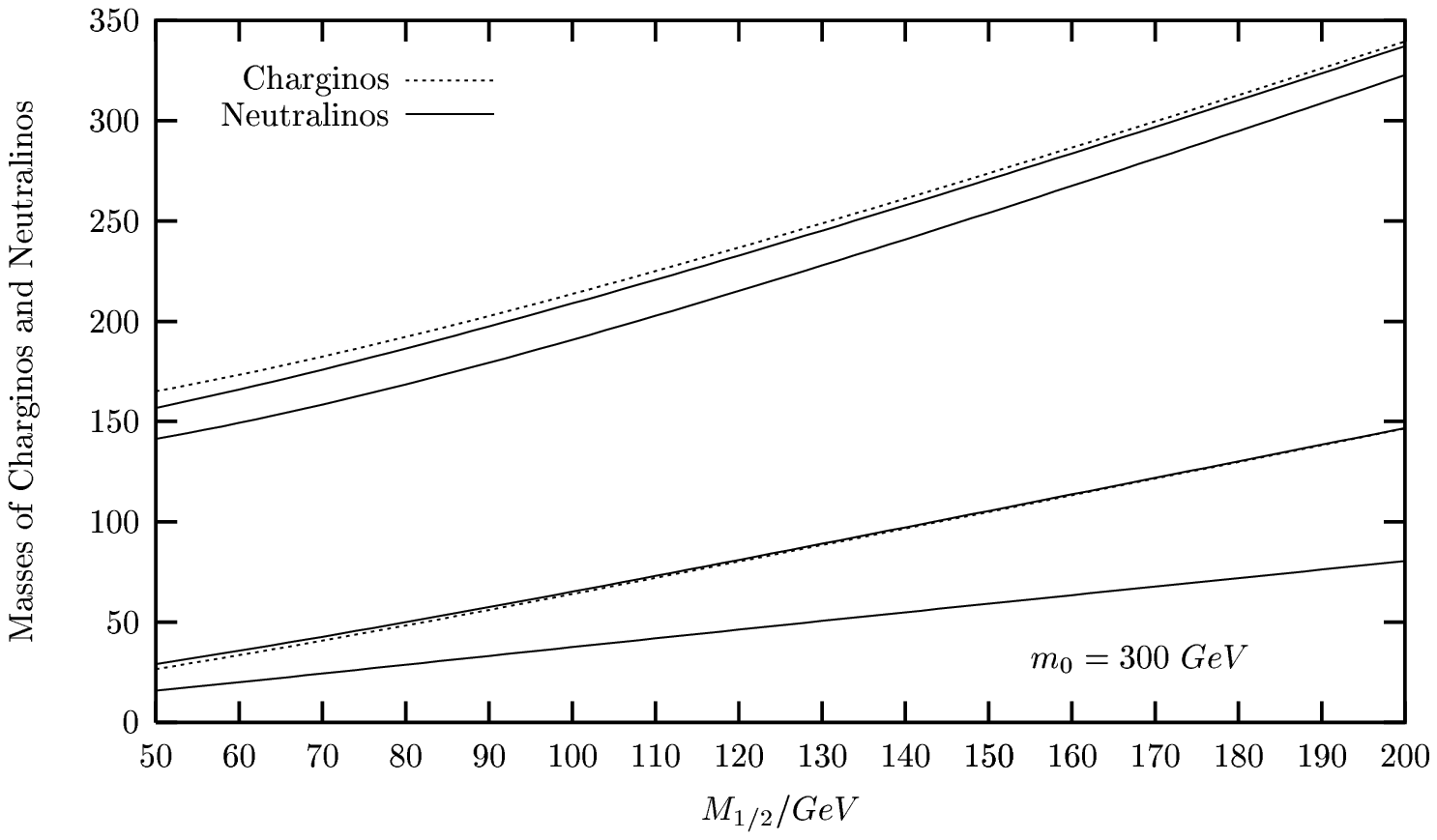}}

{\narrower\narrower\footnotesize\noindent
{\bf Figure \FigNewCharginoNeutralinoB.}
Spectrum of charginos ($m_C$) and neutralinos ($m_N$) for a range of $M_{1/2}$
and $m_0 = 300$ GeV
($A_0=0$ GeV, $M_\nu=2.0\times 10^{16}$ GeV).\par}
\bigskip}


\section{Conclusions}

The main qualitative conclusion of this study is that
LFV is not a unique prediction of SUSY GUTs, but is also
found in certain string-inspired models which do not possess a
simple gauge group. In order to illustrate this
we have calculated the minimum irreducible contributions to
LFV in a string-inspired minimal \422 model. 
The main features of this model are large \tanb\ 
and neutrino masses with an intermediate mass scale $M_\nu$.
The mechanism responsible for LFV in the 422 model is
similar to the one in \MSSM+N but here involves a much more constrained
parameter space, leading to a range of \tanb\ outside that previously
considered. Also previous studies on the
fermion mass spectrum in this model lead to a set
of well defined mixing angles which enable precise predictions of LFV
to be made. The dominant contribution
was seen to come from the amplitude $A_{R_1}$ corresponding to
sneutrinos and charginos in the loop, with the LFV controlled
by the off-diagonal contributions to the left-handed
sneutrino mass squared matrix. 
In Appendix 5 we saw that the
positive off-diagonal contribution from the F-term neutrino mass
must compete with the contribution arising from high energy RGE running
effects in the high energy region between $M_{Planck}$ and $M_{\nu}$ which is
negative and tends to cancel the F-term. The combined effect of these
two terms is largely responsible for the 
resonant suppression of the rates for \muegamma and \taumugamma
seen in Figures \FigBRmuegammaMain\ to \FigBRtaumugammaMain. 

The main quantitative conclusion is that the LFV rates
in these models are {\em substantially enhanced} compared to other models.
This conclusion is based on values of mixing angles taken from
previous studies of the fermion mass spectrum in this model.
The enhancement effect is well illustrated by the
detailed analysis of the parameter space near the current
experimental limits given in Figures \FigNewBRmuegammmaMain\ to
\FigNewCharginoNeutralinoB. In particular 
we find that the current limit on \taumugamma 
is very close to the predictions of this model, especially
for the lower values of right-handed neutrino masses which are
well motivated by the physics of neutrino masses.
If the experimental bounds on \taumugamma were improved by one order
of magnitude then this model would become severely constrained,
providing a decisive test of such models.
Since we have concentrated on the {\em minimum irreducible} amount of LFV
in the model, failure to observe \taumugamma at its predicted rate
would enable such models to be experimentally excluded.
More optimistically a direct observation of
\taumugamma could provide an indirect discovery of supersymmetry in general
and large \tanb\ string-inspired models in particular.


\newpage

\begin{center}
{\large \bf Appendix 1: Effective Theory Below $M_{PS}$.}
\end{center}

Below the 422 breaking scale $\MGUT \sim 10^{16}$ \GeV,
the model effectively reduces to the \MSSM+N model.
The effective Lagrangian is given by summing
the superpotential, scalar potential, scalar and gaugino  mass 
contributions
$\ {\cal L} = {\cal L}_W-{\cal V}-{\cal L}_m-{\cal L}_\lambda$, 
each of which we have written in the following form~:
\begin{eqnarray}
{\cal W} &=& u^c \lambda_u Q H_u + d^c \lambda_d Q H_d + \nonumber \\
         & & \nu^c \lambda_\nu L H_u + e^c \lambda_e L H_d +
             \mu H_u H_d +   
             1/2\>M_\nu \nu^c \nu^c \\
         & & \nonumber \\
{\cal V} &=& \tilde u^c \tilde\lambda_u \tilde Q H_u + 
             \tilde d^c \tilde\lambda_d \tilde Q H_d + \nonumber \\
         & & \tilde\nu^c \tilde\lambda_\nu \tilde L H_u + 
             \tilde e^c \tilde\lambda_e \tilde L H_d +
             \tilde\mu^2 H_u H_d + \hbox{h.c.} \\
         & & \nonumber \\
{\cal L}_m &=& m_{H_u}^2 \vert H_u \vert ^2 +
               m_{H_d}^2 \vert H_d \vert ^2 + \nonumber \\
           & & \tilde Q^\hc \tilde m_Q^2 \tilde Q +
               \tilde u^c \tilde m_{u^c}^2 \tilde u^{c\hc} +
               \tilde d^c \tilde m_{d^c}^2 \tilde d^{c\hc} + \nonumber \\
           & & \tilde L^\hc \tilde m_L^2 \tilde L +
               \tilde \nu^c \tilde m_{\nu^c}^2 \tilde \nu^{c\hc} +
               \tilde e^c \tilde m_{e^c}^2 \tilde e^{c\hc} \\
         & & \nonumber \\
{\cal L}_\lambda &=& 1/2 \> M_1 \, \overline{\widetilde B} \widetilde B +
                     1/2 \> M_2 \> \overline{\widetilde W}_a \widetilde W_a +
                     1/2 \> M_3 \> \overline{\widetilde G}_x \widetilde G_x
\end{eqnarray}
\noindent
which defines our conventions and notation for the soft parameters
in the low energy effective theory.


\begin{center}
{\large \bf Appendix 2: RGEs.}
\end{center}

This appendix lists the one loop RGEs which we used to run the 
parameters between $M_{GUT}$ and $M_{Planck}$ using the effective 422 model
as decribed in Section 3. 
We have negleted the wave function renormalization of the GUT Higgs
fields, consequently the equations 
resemble those of the \MSSM+N\ with effective
Yukawa couplings $\lambda_u$, $\lambda_d$, $\lambda_\nu$, $\lambda_e$,
but with the \422 gauge group instead of the standard model gauge group.

\bigskip
\noindent 
$\bullet$
The gauge group factors are given by :
$$
\vbox{
\halign{ 
# \hfill \quad &
\hfill # \hfill \quad &
\hfill # \hfill \quad &
\hfill # \hfill \quad &
\hfill # \hfill \quad &
\hfill # \hfill \quad &
\hfill # \hfill \quad &
\hfill # \hfill \quad \cr
\noalign{\smallskip\hrule\smallskip}
Group & 
i & 
$b_i$ & 
$c_i^\lambda$ & 
$c_i^h$ &
$c_i^{2h}$ & 
$c_i^F$ & 
$c_i^{F^c}$ \cr
\noalign{\smallskip\hrule\smallskip}
\noalign{\smallskip}
$SU(4)$ &
$1$ &
$-6$ &
$15/4$ & 
$0$ &
$0$ &
$15/8$ &
$15/8$ \cr
$SU(2)_L$ &
$2$ &
$1$ &
$3/2$ &
$3/4$ &
$3/2$ &
$3/4$ &
$0$\cr
$SU(2)_R$ &
$3$ &
$1$ &
$3/2$ &
$3/4$ &
$3/2$ &
$0$ &
$3/4$ \cr
\noalign{\smallskip\hrule\smallskip}
}}
$$
The $b$s displayed above account only for the contributions coming
from the $F$, $F^c$ and $h$ multiplets. More generally one can write~:
$$
(b_4,b_L,b_R)=(-6,1,1)+2n_{H_L}(1,2,0)+2n_{H_R}(1,0,2)+n_D(1,0,0)
$$
Where the second term refers to $n_{H_L}$ copies of the Higgs $H$, $H^c$
as in \refeqn{H:higgs}, the third for $n_{H_R}$ copies of GUT Higgs
$H_R$, $H_R^c$ in $(4,2,1)$, $(\bar 4,\bar 2,1)$ and the last for $n_D$
copies of $D$ sextet fields in $(6,1,1)$. These extra fields are
necessary in order to guarantee that the gauge couplings
remain unified above \MGUT \cite{king}.

\medskip
\noindent
$\bullet$
Running of gauge couplings and gauginos :
$$
16\pi^2 {d g_i \over d t} = b_i\>g_i^3 
\qquad\null\qquad
16\pi^2 {d M_i \over d t} = 2 b_i M_i\>g_i^2 
$$
$\bullet$ Running of superpotential Yukawa couplings~:
\begin{eqnarray}
16\pi^2 {d \lambda_u \over d t} 
& \!\!\!\! = & \!\!\!\!
    \lambda_u \> [ \>
3\> \tr\{\lambda_u^\hc \lambda_u\}+
    \tr\{\lambda_\nu^\hc\lambda_\nu\}+
3\lambda_u^\hc\lambda_u+
 \lambda_d^\hc\lambda_d-
  2 \> c_i^\lambda g_i^2 \> ] \nonumber \\ 
16\pi^2 {d \lambda_d \over d t}
& \!\!\!\! = & \!\!\!\!
    \lambda_d \> [ \>
3\> \tr\{\lambda_d^\hc\lambda_d\}+
    \tr\{\lambda_e^\hc\lambda_e\}+
3\lambda_d^\hc\lambda_d+
 \lambda_u^\hc\lambda_u-
  2 \> c_i^\lambda g_i^2 \> ] \nonumber \\
16\pi^2 {d \lambda_\nu \over d t} 
& \!\!\!\! = & \!\!\!\!
    \lambda_\nu \> [ \>
3\> \tr\{\lambda_u^\hc\lambda_u\}+
    \tr\{\lambda_\nu^\hc\lambda_\nu\}+
3\lambda_\nu^\hc\lambda_\nu+
 \lambda_e^\hc\lambda_e-
  2 \> c_i^\lambda g_i^2 \> ] \nonumber \\
16\pi^2 {d \lambda_e \over d t} 
& \!\!\!\! = & \!\!\!\!
    \lambda_e \> [ \>
3\> \tr\{\lambda_d^\hc\lambda_d\}+
    \tr\{\lambda_e^\hc\lambda_e\}+
3\lambda_e^\hc\lambda_e+
 \lambda_\nu^\hc\lambda_\nu-
  2 \> c_i^\lambda g_i^2 \> ] \nonumber
\end{eqnarray}
$\bullet$ Running of Higgs parameter~:
\begin{eqnarray}
16\pi^2 {d \mu \over d t}
& \!\!\!\! = & \!\!\!\!
    \mu \> [ \>
3\> \tr\{\lambda_u^\hc\lambda_u\}+
3\> \tr\{\lambda_d^\hc\lambda_d\}+
    \tr\{\lambda_\nu^\hc\lambda_\nu\}+
    \tr\{\lambda_e^\hc\lambda_e\}-
  2 \> c_i^{2h} g_i^2 \> ] \nonumber
\end{eqnarray}
$\bullet$ Running of soft triliniar Yukawa couplings~:
\begin{eqnarray}
16\pi^2 {d \tilde\lambda_u \over d t}
& \!\!\!\! = & \!\!\!\!
\phantom{2} \tilde\lambda_u \> [ \>
3\> \tr\{\lambda_u^\hc\lambda_u\}+
    \tr\{\lambda_\nu^\hc\lambda_\nu\}+
5   \lambda_u^\hc\lambda_u+
    \lambda_d^\hc\lambda_d-
2 \> c_i^\lambda g_i^2 \> ] + \nonumber \\
& \!\!\!\! \phantom{=} & \!\!\!\!
    2 \lambda_u \> [ \>
3\> \tr\{\lambda_u^\hc\tilde\lambda_u\}+
    \tr\{\lambda_\nu^\hc\tilde\lambda_\nu\}+
2   \lambda_u^\hc\tilde\lambda_u+
    \lambda_d^\hc\tilde\lambda_d+
2 \> c_i^\lambda M_i g_i^2 \> ] \nonumber \\
16\pi^2 {d \tilde\lambda_d \over d t} 
& \!\!\!\! = & \!\!\!\!
\phantom{2} \tilde\lambda_d \> [ \>
3\> \tr\{\lambda_d^\hc\lambda_d\}+
    \tr\{\lambda_e^\hc\lambda_e\}+
5   \lambda_d^\hc\lambda_d+
    \lambda_u^\hc\lambda_u-
2 \> c_i^\lambda g_i^2 \> ] + \nonumber \\
& \!\!\!\! \phantom{=} & \!\!\!\!
     2 \lambda_d \> [ \>
3\> \tr\{\lambda_d^\hc\tilde\lambda_d\}+
    \tr\{\lambda_e^\hc\tilde\lambda_e\}+
2   \lambda_d^\hc\tilde\lambda_d+
    \lambda_u^\hc\tilde\lambda_u+
2 \> c_i^\lambda M_i g_i^2 \> ] \nonumber \\
16\pi^2 {d \tilde\lambda_\nu \over d t}
& \!\!\!\! = & \!\!\!\!
\phantom{2} \tilde\lambda_\nu \> [ \>
3\> \tr\{\lambda_u^\hc\lambda_u\}+
    \tr\{\lambda_\nu^\hc\lambda_\nu\}+
5   \lambda_\nu^\hc\lambda_\nu+
    \lambda_e^\hc\lambda_e-
2 \> c_i^\lambda g_i^2 \> ] + \nonumber \\
& \!\!\!\! \phantom{=} & \!\!\!\!
    2 \lambda_\nu \> [ \>
3\> \tr\{\lambda_u^\hc\tilde\lambda_u\}+
    \tr\{\lambda_\nu^\hc\tilde\lambda_\nu\}+
2   \lambda_\nu^\hc\tilde\lambda_\nu+
    \lambda_e^\hc\tilde\lambda_e+
2 \> c_i^\lambda M_i g_i^2 \> ] \nonumber \\
16\pi^2 {d \tilde\lambda_e \over d t}
& \!\!\!\! = & \!\!\!\!
\phantom{2} \tilde\lambda_e \> [ \>
3\> \tr\{\lambda_d^\hc\lambda_d\}+
    \tr\{\lambda_e^\hc\lambda_e\}+
5   \lambda_e^\hc\lambda_e+
    \lambda_\nu^\hc\lambda_\nu-
2 \> c_i^\lambda g_i^2 \> ] + \nonumber \\
& \!\!\!\! \phantom{=} & \!\!\!\!
    2 \lambda_e \> [ \>
3\> \tr\{\lambda_d^\hc\tilde\lambda_d\}+
    \tr\{\lambda_e^\hc\tilde\lambda_e\}+
2   \lambda_e^\hc\tilde\lambda_e+
    \lambda_\nu^\hc\tilde\lambda_\nu+
2 \> c_i^\lambda M_i g_i^2 \> ] \nonumber 
\end{eqnarray}
$\bullet$ Running of soft Higgs parameter~:
\begin{eqnarray}
16\pi^2 {d \tilde\mu^2 \over d t} 
& \!\!\!\! = & \!\!\!\!
    \tilde\mu^2 \> [ \>
3\> \tr\{\lambda_u^\hc\lambda_u\}+
3   \tr\{\lambda_d^\hc\lambda_d\}+
    \tr\{\lambda_\nu^\hc\lambda_\nu\}+
    \tr\{\lambda_e^\hc\lambda_e\}-
2 \> c_i^{2h} g_i^2 \> ] + \nonumber \\
& \!\!\!\! \phantom{=} & \!\!\!\!
     2 \mu \> [ \>
3\> \tr\{\lambda_u^\hc\tilde\lambda_u\}+
3   \tr\{\lambda_d^\hc\tilde\lambda_d\}+
    \tr\{\lambda_\nu^\hc\tilde\lambda_\nu\}+
    \tr\{\lambda_e^\hc\tilde\lambda_e\}+
2 \> c_i^{2h} M_i g_i^2 \> ] \nonumber 
\end{eqnarray}
$\bullet$ Running of soft scalar masses~:
\begin{eqnarray}
16\pi^2 {d \tilde m_Q^2 \over d t} 
& \!\!\!\! = & \!\!\!\! \phantom{2} \>
 [ \> \tilde m_Q^2 \lambda_u^\hc\lambda_u+
      \lambda_u^\hc ( m_{H_u}^2 + \tilde m_{u^c}^2 ) \lambda_u+
      \tilde\lambda_u^\hc\tilde\lambda_u+ \nonumber \\
& & \!\!\!\! \phantom{2} \> \phantom{[} \> 
      \tilde m_Q^2 \lambda_d^\hc\lambda_d+
      \lambda_d^\hc ( m_{H_d}^2 + \tilde m_{d^c}^2 ) \lambda_d+
      \tilde\lambda_d^\hc\tilde\lambda_d + \hbox{h.c.} ] -
8 \> c_i^F M_i^2 g_i^2 \nonumber \\
16\pi^2 {d \tilde m_{u^c}^2 \over d t} 
& \!\!\!\! = & \!\!\!\! 2 \>
[ \> \tilde m_{u^c}^2 \lambda_u\lambda_u^\hc+
     \lambda_u ( m_{H_u}^2 + \tilde m_Q^2 ) \lambda_u^\hc+
     \tilde\lambda_u\tilde\lambda_u^\hc + \hbox{h.c.} ] -
8 \> c_i^{F^c} M_i^2 g_i^2 \nonumber \\
16\pi^2 {d \tilde m_{d^c}^2 \over d t} 
& \!\!\!\! = & \!\!\!\! 2 \>
[ \> \tilde m_{d^c}^2 \lambda_d\lambda_d^\hc+
     \lambda_d ( m_{H_d}^2 + \tilde m_Q^2 ) \lambda_d^\hc+
     \tilde\lambda_d\tilde\lambda_d^\hc + \hbox{h.c.} ] -
8 \> c_i^{F^c} M_i^2 g_i^2 \nonumber \\
16\pi^2 {d \tilde m_L^2 \over d t}
& \!\!\!\! = & \!\!\!\! \phantom{2} \>
 [ \> \tilde m_L^2 \lambda_\nu^\hc\lambda_\nu+
      \lambda_\nu^\hc ( m_{H_u}^2 + \tilde m_{\nu^c}^2 ) \lambda_\nu+
      \tilde\lambda_\nu^\hc\tilde\lambda_\nu+ \nonumber \\
& & \!\!\!\! \phantom{2} \> \phantom{[} \> 
      \tilde m_L^2 \lambda_e^\hc\lambda_e+
      \lambda_e^\hc ( m_{H_d}^2 + \tilde m_{e^c}^2 ) \lambda_e+
      \tilde\lambda_e^\hc\tilde\lambda_e + \hbox{h.c.} ] -
8 \> c_i^F M_i^2 g_i^2 \nonumber \\
16\pi^2 {d \tilde m_{\nu^c}^2 \over d t}
& \!\!\!\! = & \!\!\!\! 2 \>
[ \> \tilde m_{\nu^c}^2 \lambda_\nu\lambda_\nu^\hc+
     \lambda_\nu ( m_{H_u}^2 + \tilde m_L^2 ) \lambda_\nu^\hc+
     \tilde\lambda_\nu\tilde\lambda_\nu^\hc + \hbox{h.c.} ] -
8 \> c_i^{F^c} M_i^2 g_i^2 \nonumber \\
16\pi^2 {d \tilde m_{e^c}^2 \over d t}
& \!\!\!\! = & \!\!\!\! 2 \>
[ \> \tilde m_{e^c}^2 \lambda_e\lambda_e^\hc+
     \lambda_e ( m_{H_d}^2 + \tilde m_L^2 ) \lambda_e^\hc+
     \tilde\lambda_e\tilde\lambda_e^\hc + \hbox{h.c.} ] -
8 \> c_i^{F^c} M_i^2 g_i^2 \nonumber 
\end{eqnarray}
$\bullet$ Running of Higgs masses~:
\begin{eqnarray}
16\pi^2 {d m_{H_u}^2 \over d t} 
& \!\!\!\! = & \!\!\!\!
6 \>  \tr\{\tilde m_Q^2 \lambda_u^\hc\lambda_u+
      \lambda_u^\hc ( m_{H_u}^2 + \tilde m_{u^c}^2 ) \lambda_u+
      \tilde\lambda_u^\hc\tilde\lambda_u\}+ \nonumber \\
& & \!\!\!\!  
2 \> \tr\{\tilde m_L^2 \lambda_\nu^\hc\lambda_\nu+
     \lambda_\nu^\hc ( m_{H_u}^2 + \tilde m_{\nu^c}^2 ) \lambda_\nu+
     \tilde\lambda_\nu^\hc\tilde\lambda_\nu\}-
8 \> c_i^{h} M_i^2 g_i^2 \nonumber \\
16\pi^2 {d m_{H_d}^2 \over d t}
& \!\!\!\! = & \!\!\!\!
6 \> \tr\{\tilde m_Q^2 \lambda_d^\hc\lambda_d+
     \lambda_d^\hc ( m_{H_d}^2 + \tilde m_{d^c}^2 ) \lambda_d+
     \tilde\lambda_d^\hc\tilde\lambda_d\}+ \nonumber \\
& & \!\!\!\!
2 \> \tr\{\tilde m_L^2 \lambda_e^\hc\lambda_e+
     \lambda_e^\hc ( m_{H_d}^2 + \tilde m_{e^c}^2 ) \lambda_e+
     \tilde\lambda_e^\hc\tilde\lambda_e\}-
8 \> c_i^{h} M_i^2 g_i^2 \nonumber
\end{eqnarray}


\begin{center}
{\large \bf Appendix 3: Diagonalisation of Mass Matrices.}
\end{center}

The soft trilinear Yukawa couplings and mass terms of the 422 theory are  
$\tilde F^c\tilde\lambda\>\tilde F h$ and
$\tilde F^\hc \tilde m_F^2 \tilde F$,
$\tilde F^c \tilde m_{F^c}^2 \tilde F^{c\hc}$.
Below the symmetry breaking scale
$\tilde m^2_F$ 
splits into $\tilde m^2_Q$, $\tilde m^2_L$, and $\tilde m^2_{F^c}$
splits into $\tilde m^2_{u^c}$, 
$\tilde m^2_{d^c}$,$\tilde m^2_{e^c}$, $\tilde m^2_{\nu^c}$~.
\noindent
We now specify how 
Yukawa and soft scalar mass matrices are diagonalised~:
\begin{equation}
S^u \lambda_u T^{u\hc} = \lambda_u(d)\qquad
S^d \lambda_d T^{d\hc} = \lambda_d(d)
\end{equation}
\begin{equation}
S^\nu \lambda_\nu T^{\nu\hc} = \lambda_\nu(d)\qquad 
S^e \lambda_e T^{e\hc} = \lambda_e(d) 
\end{equation}
\begin{equation}
\tilde T^Q \tilde m_Q^2 \tilde T^{Q\hc} = \tilde m_Q^2(d)\qquad
\tilde T^L \tilde m_L^2 \tilde T^{L\hc} = \tilde m_L^2(d)
\end{equation}
\begin{equation}
\tilde S^{u^c} \tilde m_{u^c}^2 \tilde S^{u^c\hc} = \tilde m_{u^c}^2(d)\qquad
\tilde S^{d^c} \tilde m_{d^c}^2 \tilde S^{d^c\hc} = \tilde m_{d^c}^2(d)
\end{equation}
\begin{equation}
\tilde S^{\nu^c} \tilde m_{\nu^c}^2 \tilde S^{\nu^c\hc} 
= \tilde m_{\nu^c}^2(d)\qquad
\tilde S^{e^c} \tilde m_{e^c}^2 \tilde S^{e^c\hc} = \tilde m_{e^c}^2(d)
\end{equation}
\noindent
The left-handed neutrinos obtain a small mass $\sim m^2_\nu / 4 M_\nu$
after diagonalization of :
\begin{equation}
{\cal L} = - (\nu \quad \nu^c)
             \left( \matrix{ 0           & 1/2\>m_\nu^\tp \cr
                             1/2\>m_\nu & M_\nu         \cr} \right)
             \left( \matrix{ \nu \cr \nu^c \cr} \right) + \hbox{h.c.}
\end{equation}
We introduce $\tilde S^n$, $\tilde S^l$ which
diagonalise the $6\times 6$ sneutrino $\tilde M^{n2}$, 
selectron $\tilde M^{l2}$ mass matrices~:
\begin{equation}
\matrix{
\tilde S^{n} \tilde M^{n2} \tilde S^{n\hc}=\tilde M^{n2}(d) &
\tilde M^{n2}= \left( 
               \matrix{ \tilde M_{LL}^{n2} & \tilde M_{LR}^{n2} \cr
                        \tilde M_{RL}^{n2} & \tilde M_{RR}^{n2} \cr} 
               \right) &
\tilde S^{n}= \left( 
              \matrix{ \tilde S_{LL}^n & \tilde S_{LR}^n \cr
                       \tilde S_{RL}^n & \tilde S_{RR}^n \cr} 
              \right) \cr
& & \cr
\tilde S^{l} \tilde M^{l2} \tilde S^{l\hc}=\tilde M^{l2}(d) &
\tilde M^{l2}= \left( 
               \matrix{ \tilde M_{LL}^{l2} & \tilde M_{LR}^{l2} \cr
                        \tilde M_{RL}^{l2} & \tilde M_{RR}^{l2} \cr} 
               \right) &
\tilde S^{l}= \left( 
              \matrix{ \tilde S_{LL}^l & \tilde S_{LR}^l \cr
                       \tilde S_{RL}^l & \tilde S_{RR}^l \cr} 
              \right) \cr}
\end{equation}
\noindent
The left-handed sneutrino mass matrix is given by~:
\footnote{Note that the neutrino Dirac masses contribute to the
left-handed sneutrino masses even though the right-handed (singlet)
neutrino superfield is very heavy $\sim M_{\nu}$.
Such terms arise from the F-terms $F_{\nu^c}$ which do not
involve the heavy $\nu^c$ field.
Also note that, since $m_\nu = v_u \lambda_\nu$, 
we have $m_{\nu_\tau} \sim m_{top}$, therefore for low $m_0$, $M_{1/2}$
it is not true that $\tilde m_L \gg m_\nu$. 
Furthermore in expression \refeqn{leftleftsneutrinomass}
we have negleted terms of order $m_\nu / M_\nu$ which is perfectly
valid (The same however does not apply to the analogous contribution
to the mass of left-handed neutrinos, since they are made massive
exactly because of left-right mixing through $\lambda_\nu$).}
\begin{equation}
\tilde M^{n2}_{LL} 
= {\tilde m}_L^2+ m_\nu^\hc m_\nu + m_Z^2 Z_{\nu_L} c_{2\beta}
\label{leftleftsneutrinomass}
\end{equation}
\noindent

We denote the eigenvalues of 
$\tilde M^{n2}$ and $\tilde M^{l2}$ by
$\tilde m^2_n$~($\tilde m^2_{n_A}=\tilde m^2_{n_{A_L}}$,
$\tilde m^2_{n_{\dot A}}=\tilde m^2_{n_{A_R}}$) and
$\tilde m^2_l$~($\tilde m^2_{l_A}=\tilde m^2_{l_{A_L}}$, 
$\tilde m^2_{l_{\dot A}}=\tilde m^2_{l_{A_R}}$) respectively.
\noindent
The $\tilde U$s are given by :

\begin{equation}
\tilde U^{n,l} = \left(
                 \matrix{\tilde U_{LL}^{n,l} & \tilde U_{LR}^{n,l} \cr
                         \tilde U_{RL}^{n,l} & \tilde U_{RR}^{n,l} \cr}
                         \right)
=                        \left(
     \matrix{\tilde S_{LL}^{n,l} T^{e\hc} & \tilde S_{LR}^{n,l} S^{e\hc} \cr
       \tilde S_{RL}^{n,l} T^{e\hc} & \tilde S_{RR}^{n,l} S^{e\hc} \cr}
                         \right)
\end{equation}

\noindent
Finally we provide the expressions for $J$ and $H$ :
\begin{equation}
J_{ijA} =  \sum_{k=1,2}
           (g_1 S_{ik}^{C\hc})(g_1 T_{kj}^C) \>
           m_{C_k} / \tilde m_{n_A}^2 \>
           J_{kA}
\qquad
J_{kA} = {\cal J}(m_{C_k}^2/\tilde m_{n_A}^2)
\end{equation}
\begin{equation}
H_{pqA} =   \sum_{r=1..4}
            (g_p S_{pr}^{N\hc})(g_q S_{rq}^N) \>
            m_{N_r} / \tilde m_{l_A}^2 \>
            H_{rA}
\qquad
H_{rA} = {\cal H}(m_{N_r}^2/\tilde m_{l_A}^2)
\end{equation}
\noindent
Here $g_{p,q}=(g', g, g, g)$ and the function $\cal J$ ($\cal H$) arises
from chargino (neutralino) loop integration \cite{hisano}.
The (supersymmetric state) indices $i$, $j$, $p$ and $q$ can take
values among :
$i=(1,2)=(\tilde W_R^-,\>\tilde H_R^-)$,
$j=(1,2)=(\tilde W_L^-,\>\tilde H_L^-)$,
$p$ and $q=(1,2,3,4)=(\tilde B,\>\tilde W^0,\>\tilde H_d^0,\>\tilde H_u^0)$.  
$S^C$ and $T^C$ diagonalise ($S^C M^C T^{C\hc}=M^C(d)$) the $2\times 2$ 
chargino mass matrix $M^C$  which has
eigenvalues $m_{C_k}$. 
Similarly $S^N$ diagonalises ($S^N M^N S^{N\hc}=M^N(d)$) the $4\times 4$
neutralino mass matrix $M^N$  which has
eigenvalues $m_{N_r}$.
\begin{eqnarray}
M^C &=& 
\left(\matrix{M_2 & \sqrt 2\>m_W s_\beta \cr
              \sqrt 2\>m_W c_\beta & \mu \cr}\right) \\
& & \nonumber \\
M^N &=& 
\left(\matrix{ M_1 & 0 & -m_Z c_\beta s_\theta &  m_Z s_\beta s_\theta \cr
               0 & M_2 &  m_Z c_\beta c_\theta & -m_Z s_\beta c_\theta \cr
               -m_Z c_\beta s_\theta &  m_Z c_\beta c_\theta & 0 & -\mu \cr
                m_Z s_\beta s_\theta & -m_Z s_\beta c_\theta & -\mu & 0 \cr}
      \right)
\end{eqnarray}


\begin{center}
{\large \bf Appendix 4: Fermion Masses and Mixing Angles.}
\end{center}

This appendix is intended to give an overview of fermion masses and
mixing angles predicted by this model. Ultimately we have in mind their
comparison with the available data. We will focus on the effects
associated with the running of the parameters between the weak and GUT
scale and their variation with the right-handed neutrino decoupling scale
$M_\nu$. For convenience we show below the CKM matrix \cite{pdb}~:
\begin{eqnarray}
V^Q = \left( \matrix{ 0.9747-0.9759 &  0.218-0.224  &  0.002-0.007  \cr
                       0.218-0.224  & 0.9735-0.9751 &  0.032-0.054  \cr
                       0.003-0.018  &  0.030-0.054  & 0.9985-0.9995 \cr}
                      \right)
\end{eqnarray}

The results in Tables 1-4 correspond to input values of $\alpha_s=0.115$
and $m_{bottom}=4.25$ GeV. Variations due to different choices of
these parameters are significant
\footnote{See for example Figure \FigLambdaTop.}
and have been partially considered in \cite{ben&king456}. 
Since we worked with one-loop RGE all the tables are independent of the
Planck scale parameters $m_o$, $M_{1/2}$ and $A_0$.

\bigskip
\begin{center}
\begin{tabular}{|c|l|l|l|l|l|l|} \hline

\multicolumn{7}{|c|}{Table 1} \\ \hline

$M_\nu / M_{GUT} $ & 
$Y^1$ & $Y^{Ad}$ & $Y^B$ & $Y^C$ & $Y^D$ & $Y^{33}$ \\ \hline
1.0000 & 0.0110 & -0.00032 & -0.0151 & 0.0056 & 0.0074 & 0.958 \\ 
0.0001 & 0.0131 & -0.00038 & -0.0194 & 0.0067 & 0.0087 & 1.310 \\ \hline

\end{tabular}
\end{center}
\bigskip

In Table 1, we show the values that the $Y$ operators appearing in
equations \refeqn{up:opr}-\refeqn{nu:opr}
take at the GUT scale, for two values of $M_\nu=M_{GUT}$ and
$M_\nu=2\times 10^{12}$. Operators B and 33 can be seen to be the most
sensitive to $M_\nu$.

\bigskip
\begin{center}
\begin{tabular}{|c|l|l|l|l|} \hline

Table 2 & 
\multicolumn{2}{c|}{$V^Q$}  &
\multicolumn{2}{c|}{$V^L$} \\ \cline{2-3} \cline{3-5} 

     &
Weak &
GUT  &
Weak &
GUT  \\ \hline

$V_{12}$ &
0.2210   &
0.2210   &
0.0625   &
0.0625   \\ 

$V_{11}$ &
0.9752   &
0.9752   &
0.9980   &
0.9980   \\ 

$V_{22}$ &
0.9743   &
0.9747   &
0.9974   &
0.9976   \\

$V_{33}$ &
0.9990   &
0.9995   &
0.9993   &
0.9995   \\ \hline

\end{tabular}
\end{center}
\bigskip

In Table 2, we have collected the mixing angles which are approximately
insensitive to changes in $M_\nu$ and stable relative to RGE effects. 
We denoted the CKM matrix by $V^Q$ and the leptonic counterpart 
by $V^L$ ($V_{21} \sim V_{12}$). The Clebsh factors in equations 
\refeqn{up:opr}-\refeqn{nu:opr} imply $V^L_{12} \sim V^Q_{12}/4$.
\footnote{It is relevant to note that the $Y$ operators
were chosen because they can, not only account for the 
experimental fermion mass
pattern but predict successfully `natural' mixing angles as well. By this
we mean that we were careful to select them in such a way that none
arises as the residue of an almost complete cancellation of the
contributions coming from the up and down Yukawa matrices.}

\bigskip
\begin{center}
\begin{tabular}{|c|l|l|l|l|} \hline

\multicolumn{5}{|c|}{Table 3} \\ \hline

$M_\nu = M_{GUT}$ &
\multicolumn{2}{c|}{Weak} &
\multicolumn{2}{c|}{GUT} \\ \hline

$V^Q_{AB}$ &
Input &
Output &
Input &
Output \\ \hline

$V^Q_{23}$ &
0.0430 &
0.0430 &
0.0310 &
0.0310 \\

$V^Q_{32}$ &
0.0429 &
0.0437 &
0.0309 &
0.0315 \\ 

$V^Q_{13}$ &
0.0045 &
0.0078 &
0.0032 &
0.0056 \\

$V^Q_{31}$ &
0.0051 &
0.0019 &
0.0036 &
0.0013 \\ \hline

$V^L_{AB}$ &
&
&
& \\ \cline{1-1}

$V^L_{23}$ &
&
0.0352 &
&
0.0315 \\ 

$V^L_{32}$ &
Unknown&
0.0352 &
Unknown&
0.0315 \\ 

$V^L_{13}$ &
&
0.0016 &
&
0.0014 \\

$V^L_{31}$ &
&
0.0006 &
&
0.0005 \\ \hline

\end{tabular}
 
\end{center}
\bigskip

In Table 3 we include the remaining $V^Q$ and $V^L$ entries not
present in Table 2 ($M_\nu = 2\times 10^{16}$). 
The reason for the discrimination is threefold.
Firstly because we are now confronted with values that are more sensitive to
variations in $M_\nu$. For example, taking $M_\nu = 10^{-4} \times M_{GUT}$
effects the values shown to about 7 \%. Secondly because the mixings in
Table 3 are generally not as stable to RGE effects as the ones in
Table 2. And finally because we wanted to call attention to the fact
that the values that are actually used when we computed LFV processes,
denoted by $\langle$Output$\rangle$, are not exactly the same as
the ones we have available from experiment $\langle$Input$\rangle$. 
The discrepancy arises when we replace the GUT Yukawa 
couplings by others parameterized by our set of operators arranged
in a successful `Texture'. 

\bigskip
\begin{center}
\begin{tabular}{|l|l|l|l|} \hline

Table 4 &
$m_d$ &
$m_s$ &
$m_t$ \\ 

&
7.8 MeV &
214 MeV &
178-175 GeV \\ \hline

$m_{\nu_e}$ &
$m_{\nu_\mu}$ &
$m_{\nu_\tau}$ &
$m^2_{\nu_\tau}/4M_\nu$ \\

0.2 MeV &
760 MeV &
115-122 GeV &
1.6-$10^{-4}$ eV \\ \hline

\end{tabular}
\end{center}
\bigskip

Finally in Table 4, we present the predictions for some
fermion masses. Whenever two values are shown for the same
parameter, the first is associated with $M_\nu = 10^{-4} M_{GUT}$
while the second with $M_\nu = M_{GUT}$.
The down (strange) quark has a mass within the $5-15$
($100-300$) MeV range quoted in \cite{pdb}. 
The $m_{\nu}$ values correspond to the unphysical mass directly 
obtained from the neutrino Yukawa couplings 
(for example $m_{\nu_\tau} \sim v_u (\lambda_\nu)_{33}$).
On the other hand, the physical mass of the tau-neutrino is correctly
obtained after taking into account the see-saw suppression mechanism
which forces it to scale as $m^2_{\nu_\tau}/ 4M_\nu$. 
In all cases we obtained predictions fairly compatible with
experimental data.


\begin{center}
{\large \bf Appendix 5: Analysis of Suppression in LFV Decays.}
\end{center}

In this appendix we investigate the origin of the suppression in the
LFV decays observed in Figures \FigBRmuegammaMain-\FigBRtaumugammaMain. 
In section 4.1 we showed that the amplitude $A_{R_1}$ gave the dominant
contribution to the LFV branching ratios (see Figure 4.)
Furthermore as shown in Eqs.\refeqn{12}, \refeqn{23} the LFV due to $A_{R_1}$
is controlled by the off-diagonal elements of the matrices
$\tilde{U}^n_{LL}$ which are involved in the diagonalisation
of the left-handed sneutrino mass squared matrix
(see Appendix 3.)
The mass matrix for left-handed 
sneutrinos  
was given in equation \refeqn{leftleftsneutrinomass}~:
\begin{equation}
\tilde M^{n2}_{LL} 
= {\tilde m}_L^2+ m_\nu^\hc m_\nu + m_Z^2 Z_{\nu_L} c_{2\beta}
\label{app5:1}
\end{equation}
In the basis in which charged leptons are diagonal we write
\begin{equation}
T^e \tilde M^{n2}_{LL} T^{e\hc} = 
(T^e \tilde T^{L\hc}) {\tilde m}^2_L(d) (\tilde T^L T^{e\hc})+
(T^e T^{\nu\hc}) m_\nu^2(d) (T^\nu T^{e\hc})+\ldots
\label{sneutrinoeqn}
\end{equation}
and one immediately recognises the SUSY $T^e \tilde T^{L\hc}$ and non-SUSY
$T^e T^{\nu\hc}$ mixing matrices reflecting the mismatch between the
scalar-fermion (superpartners) and charged-neutral lepton
eigen-mass basis respectively, which leads to LFV.
As we shall see the
LFV contributions coming from ${\tilde m}_L^2$ and
$m_\nu^2$ in \refeqn{sneutrinoeqn} add destructively and it is this effect
that, for some small regions of the available parameter space, 
leads to LFV being resonantly suppressed.

Now we would like to make some simple analytic estimate of
this effect, which we shall do by examining the off-diagonal
contributions to $\tilde  M^{n2}_{LL}$ which are responsible for
flavour-violation.
We start by noting that the last term in Eq.\refeqn{app5:1}, 
being universal, is not
relevant to what follows, therefore we concentrate on
$\tilde m^2_L$ and $m_\nu$ which are responsible for off-diagonal 
flavour-violating
entries in $\tilde  M^{n2}_{LL}$. Since $m_\nu = v_u \lambda_\nu$ we
can see that its presence in \refeqn{app5:1} cannot be neglected whenever 
$\tilde m^2_L$ is driven by low $m_0$, $M_{1/2}$. With the help of
Appendix 2, we solve approximately the RGE for $\tilde m^2_L$ :
\begin{equation}
\tilde m^2_L \sim m_0^2-\Delta E\>[\>(6\>m^2_0+A^2_0)
(\lambda_\nu^\hc \lambda_\nu + \lambda_e^\hc \lambda_e)
-8\>c\>M_{1/2}^2\>g^2\>]
\label{app5:2}
\end{equation}
\noindent
where $\Delta E = \ln( M_{Planck} / M_\nu ) / 16 \pi^2$.
In order to illustrate the origin of the suppression in LFV processes
in this model, it is convenient to adopt a set of simplifying
assumptions which will make the argument clearer. We start by fixing
$A_0 = 0$ and $M_\nu = M_{GUT}$, therefore concentrating only on the
Planck scale $m_0$, $M_{1/2}$ parameters. Additionally
it is useful to consider a basis in which the $\lambda_e$ coupling is
diagonal. This means that all LFV will be accounted by off-diagonal 
$\lambda_\nu$ entries. Finally we introduce a particularly relevant
second order effect in \refeqn{app5:2} which changes 
$6\>m_0^2 \to 6\>m_0^2+8\>c\>M_{1/2}^2\>\Delta E\>g^2$.
Hence, we arrive at a more transparent form for \refeqn{app5:1} which has
off-diagonal values given by :
\begin{equation}
\tilde M^{n2}_{LL}|_{off-diag}
 = [\>v_u^2 - \Delta E\> 
                     ( 6\>m_0^2+8\>c\>M_{1/2}^2\>\Delta E\>g^2)] \> 
                     \lambda_\nu^\hc \lambda_\nu
\label{app5:3}
\end{equation}
The essential point is that the off-diagonal
terms in the matrix contain two contributions: the
first from the Dirac neutrino mass matrix squared (arising from the
$F_{\nu^c}$-term contribution to the potential) which contributes
{\em positively}, and the second from the RGE running of the soft mass $m_0^2$
due to Yukawa corrections in the high energy region between 
$M_{Planck}$ and $M_{\nu}$
which induce off-diagonal {\em negative} contributions.
Although this expression was obtained in a rudimentary way it can
nevertheless account for some qualitative features displayed in
Figures \FigNewBRmuegammmaMain\ and \FigNewBRtaumugammmaMain. 
Firstly it sets two distinct possible
regions depending on $v_u^2$ being bigger/smaller than $m_0^2$,
$M_{1/2}^2$ corresponding to positive/negative values of 
$(\tilde M^{n2}_{LL})_{AB}$ respectively ($A \ne B$). 
Secondly it confirms that the scale $\bar m_0^2$ at 
which \refeqn{app5:3} vanishes,
decreases for increasing $M_{1/2}$. Finally it shows that $\bar m_0^2$
is the same for all types of LFV decays, ie independent of the initial
and final families involved. However the (over) simplification
of the approximations in \refeqn{app5:3} means that this equation
cannot be used to reliably estimate $\bar m_0^2$ or to check the higher
sensitivity to $M_{1/2}$ for the tau relative to the muon decay.



\begin{thebibliography}{99}

\bibitem{bh}
R.Barbieri, L.J.Hall, 
Phys. Lett. {\bf B338}, 212 (1994).

\bibitem{barb&hall}
R.Barbieri, L.J.Hall, 
Nucl.Phys. {\bf B338}, 212 (1994).

\bibitem{SUSYLFV}
F.Borzumati and A.Masiero,
Phys. Rev. Lett. {\bf 57}, 961 (1986);\\
L.J. Hall, V.A. Kostelecky and S.Raby,
Nucl. Phys. {\bf B267}, 415 (1986);\\
T.Kosmas, G.K.Leontaris and J.D.Vergados,
Phys. Lett. {\bf B219}, 457 (1989).

\bibitem{SUGRA}
E.Cremmer, S.Ferrara and J.Scherk,
Phys. Lett. {\bf B74}, 61 (1978); \par
H.P.Nilles,
Phys.Rep. {\bf 110}, 1 (1984).

\bibitem{review}
V.Barger, M.S.Berger, P.Ohmann, R.J.N.Philips,
\texttt{hep-ph/9308233}.

P.Langacker, N.Polonsky,
Phys.Rev. {\bf D49}, 1454 (1994)

G.Kane, C.Kolda, L.Roszkowski, J.Wells,
Phys.Rev. {\bf D49}, 6173 (1994)

N.Polonsky, A.Pomarol,
Phys.Rev {\bf D51}, 6532 (1995) 


\bibitem{hisano}
J.Hisano, T.Moroi, K.Tobe, M.Yamaguchi, 
Phys.Rev. {\bf D53}, 2442 (1996)

\bibitem{shafi}
S. F. King and Q. Shafi, 
\texttt{hep-ph/9711288}, \texttt{CERN-TH/97-314}, \par
(to appear in Physics Letters B.)

\bibitem{pati}
J.C.Pati, A.Salam, 
Phys.Rev. {\bf D10}, 275 (1974).

\bibitem{antoniadis}
I.Antoniadis, G.K.Leontaris, 
Phys.Lett. {\bf B216},  333 (1989); \par
I. Antoniadis, G. K. Leontaris and J. Rizos, 
Phys. Lett. {\bf B245}, 161 (1990) .

\bibitem{ben&king459}
B.C.Allanach, S.F.King,
Nucl.Phys {\bf B459}, 75 (1996).

\bibitem{arkani}
N.Arkani-Hamed, H.Cheng, L.J.Hall, 
\texttt{hep-ph/9508288}.

\bibitem{softleo}
G.K. Leontaris, N.D. Tracas
\texttt{NTUA-65-97A}, \texttt{hep-ph/9709510}; \par
G.K. Leontaris, N.D. Tracas
\texttt{NTUA-70/78}, \texttt{hep-ph/9803320}. 

\bibitem{arason&castano} 
H.Arason, D.J.Castano, E.J.Piard, P.Ramond, 
Phys.Rev. {\bf D47}, 232 (1993).

\bibitem{bbo}
V.Barger, M.S.Berger, P.Ohmann, 
Phys.Rev. {\bf D47}, 1093 (1993).

\bibitem{king}
S. F. King,  Phys. Lett. {\bf B 325}(1994) 129.

\bibitem{anderson}
G.Anderson, S.Dimopoulos, L.J.Hall, S.Raby, G.D.Starkman, 
Phys.Rev. {\bf D49}, 3660 (1994).

\bibitem{texture}
H.Georgi, C.Jarlskog, 
Phys.Lett. {\bf B89}, 297 (1978).

P.Ramond, R.G.Roberts, G.G.Ross,
Nucl.Phys. {\bf B406}, 19 (1993)

L.J.Hall, A.Rasin,
Phys.Lett. {\bf B315}, 164 (1993)

\bibitem{ben&king456}
B.C.Allanach, S.F.King, 
Nucl.Phys {\bf B456}, 57 (1995).

\bibitem{ben&king353}
B.C.Allanach, S.F.King, 
Phys.Lett. {\bf B353}, 477 (1995).

\bibitem{martin&vaughn}
S.P.Martin, M.T.Vaughn, 
\texttt{hep-ph/9311340}.

\bibitem{ben&king328}
B.C.Allanach, S.F.King, 
Phys.Lett. {\bf B328}, 360 (1994).

\bibitem{gorshny&tarasov}
S.G.Gorshny, A.L.Kataev, S.A.Larin, 
Phys.Lett. {\bf B135}, 457 (1984).

O.V.Tarasov, A.A.Vladimirov, A.Y.Zharkov, 
Phys.Lett. {\bf B93}, 429 (1980).

\bibitem{strumia}
R.Barbieri, L.J.Hall, A.Strumia, 
Nucl.Phys. {\bf B445}, 219 (1995).

\bibitem{pdb}
Particle Data Book, Phys.Rev. {\bf D54}, 58 (1996)

\bibitem{ALEPH}
ALEPH Collaboration (R. Barate \etal), 
Report No. \texttt{CERN--PPE--97--128} (\texttt{hep--ex/9710012}), 
Sep. 1997, submitted to {\sl Z. Phys. C}.

\bibitem{DELPHI}
DELPHI Collaboration (P. Abreu \etal), {\sl Eur. Phys. J.} {\bf C1},
1 (1998).

\bibitem{L3}
L3 Collaboration (M. Acciarri \etal), 
Report No. \texttt{CERN--PPE--97--130}, \par
Sep. 1997, submitted to {\sl Phys. Lett. B}.

\bibitem{OPAL}
OPAL Collaboration (K. Ackerstaff \etal), 
Report No. \texttt{CERN--PPE--97--083}, (\texttt{hep--ex/9708018}), 
submitted to {\sl Z. Phys. C}.

\end{thebibliography}
\end{document}